\documentclass{aa}  

\usepackage{graphicx}
\usepackage{txfonts}
\usepackage{color}
\usepackage{natbib}
\usepackage[switch]{lineno}
\usepackage{tabularx}
\bibpunct{(}{)}{;}{a}{}{,}
\begin{document} 

   \title{Spatially resolved origin of mm-wave linear polarization\\
   in the nuclear region of 3C\,84}

   \author{
   J. -Y. Kim\inst{1}
   \and T. P. Krichbaum\inst{1}
   \and A. P. Marscher\inst{2}
   \and S. G. Jorstad\inst{2,3}
   \and I. Agudo\inst{4}
   \and C. Thum\inst{5}
   \and J. A. Hodgson\inst{6}
   \and N. R. MacDonald\inst{1}
   \and E. Ros\inst{1}
   \and R. -S. Lu\inst{1,7}
   \and M. Bremer\inst{8} 
   \and P. de Vicente\inst{9}
   \and M. Lindqvist\inst{10} 
   \and S. Trippe\inst{11}
   \and J. A. Zensus\inst{1}
          }

   \institute{
   Max-Planck Institut f\"ur Radioastronomie, Auf dem H\"ugel 69, 53121 Bonn, Germany %
   \\         \email{jykim@mpifr-bonn.mpg.de} 
   \and       Institute for Astrophysical Research, Boston University, 725 Commonwealth Avenue, Boston, MA 022, USA %
   \and       Astronomical Institute, St. Petersburg State University, Universitetskij Pr. 28, Petrodvorets, 198504 St. Petersburg, Russia %
   \and       Instituto de Astrof\'{i}sica de Andaluc\'{i}a (CSIC), Apartado 3004, 18080, Granada, Spain %
   \and       Instituto de Radio Astronomía Millim\'{e}trica, Avenida Divina Pastora, 7, Local 20, E–18012 Granada, Spain %
   \and       Korea Astronomy and Space Science Institute, 776 Daedeokdae-ro, Yuseong-gu, Daejeon, 30455, Korea %
   \and       Shanghai Astronomical Observatory, Chinese Academy of Sciences, 80 Nandan Road, Shanghai 200030, China %
   \and       Institut de Radio Astronomie Millim\'{e}trique, 300 rue de la Piscine, Domaine Universitaire, Saint Martin d'H\`eres 38406, France %
   \and       Observatorio de Yebes (IGN), Apartado 148, E-19180 Yebes, Spain %
   \and       Department of Space, Earth and Environment, Chalmers University of Technology, Onsala Space Observatory, 439 92 Onsala, Sweden %
   \and       Department of Physics and Astronomy, Seoul National University, 1 Gwanak-ro, Gwanak-gu, Seoul 08826, Korea %
             }

   \date{Received --; accepted --}

 
  \abstract
   {
   We report results from a deep polarization imaging of the nearby radio galaxy 3C\,84 (NGC\,1275).
   The source was observed with the Global Millimeter VLBI Array (GMVA) at 86\,GHz 
   at an ultra-high angular resolution of $50\mu$as (corresponding to 250$R_{s}$).
   We also add complementary multi-wavelength data from the
   Very Long Baseline Array (VLBA; 15 \& 43\,GHz) and from 
   the Atacama Large Millimeter/submillimeter Array (ALMA; 97.5, 233.0, and 343.5\,GHz).
    At 86\,GHz, we measure a fractional linear polarization of $\sim2$\% in the VLBI core region. 
    The polarization morphology suggests that the emission is associated with an underlying limb-brightened jet.
    The fractional linear polarization is lower at 43 and 15\,GHz ($\sim0.3-0.7$\% and $<0.1$\%, respectively).
    This suggests an increasing linear polarization degree towards shorter wavelengths on VLBI scales.
    We also obtain a large rotation measure (RM) of $\sim10^{5-6}~{\rm rad/m^{2}}$ in the core 
    at $\gtrsim$43\,GHz.
    Moreover, the VLBA 43\,GHz observations show a variable RM in the VLBI core region 
    during a small flare in 2015.
    Faraday depolarization and Faraday conversion in an inhomogeneous and mildly relativistic plasma 
    could explain the observed linear polarization characteristics and the previously measured
    frequency dependence of the circular polarization.
    Our Faraday depolarization modeling suggests that
    the RM most likely originates from an external screen with a highly uniform RM distribution.
    To explain the large RM value, the uniform RM distribution, and the RM variability, 
    we suggest that the Faraday rotation is caused by a boundary layer in a transversely stratified jet.
    Based on the RM and the synchrotron spectrum of the core, 
    we provide an estimate for the magnetic field strength and the electron density of the jet plasma.
    }

   \keywords{
   Galaxies: active -- Galaxies: jets -- Galaxies: individual (NGC\,1275, 3C\,84) -- Techniques: interferometric, polarimetric
               }

   \titlerunning{GMVA 3C\,84 polarimetry}
   \authorrunning{J.-Y. Kim et al.}
               
   \maketitle
%
%


\section{Introduction}\label{sec:intro}

Magnetic fields in the vicinity of central black holes (BHs) could play a significant role
in the formation, collimation, and acceleration of relativistic jets in active galactic nuclei (AGN)
(e.g., \citealt{bz77,bp82}).
Polarization sensitive  very long baseline interferometry (VLBI) 
is an important technique to study the topology and strength of the magnetic fields at the base of AGN jets.
Synchrotron theory predicts linear polarization of up to $m_{L}\sim70$\% for optically thin emission and well ordered magnetic 
fields \citep{pacholczyk}. However, VLBI observations typically find much lower linear polarizations ($m_{L} \lesssim5$\%) in the VLBI cores of AGN-jets \citep{lister05,jorstad07}.
The origin of the weak linear polarization may be due to
(i) high synchrotron self-absorption opacity,
(ii) a large Faraday depth (e.g., \citealt{taylor06}),
and/or
(iii) highly disordered magnetic fields within the observing beam, e.g., induced by a turbulent plasma (e.g., \citealt{marscher14}).
The observed linear polarization can be also reduced by bandwidth depolarization, if Faraday rotation is large.
Observations at millimeter wavelengths help to overcome the opacity barrier and are less affected by Faraday depolarization than in the cm-bands. 
Furthermore, owing to its small beam size, VLBI observations at millimeter wavelengths (mm-VLBI) 
have a small in-beam depolarization, in comparison to single-dish or connected interferometer array observations. 
Thus mm-VLBI polarimetry is a suitable technique to observe and study the polarization properties of compact radio sources
(see also \citealt{boccardi17}).

The radio galaxy 3C\,84 (Perseus A, NGC\,1275) is an intriguing source
whose linear polarization in the core region at cm-wavelengths is extremely weak ($<0.1$\% at $\leq$15\,GHz; \citealt{taylor06}).
This is much lower than typically observed in other AGN jets \citep{lister05}.
3C\,84 is also a relatively nearby source at a redshift of $z=0.0176$
and a luminosity distance of $d_{L}=75$~Mpc \citep{3c84_dist}\footnote{
We assume $\Lambda$CDM cosmology with $H_{0}=71$km/s/Mpc, $\Omega_{\Lambda}=0.73$ and $\Omega_{M}=0.27$.
}.
The mass of its central supermassive black hole (SMBH) has been estimated to be $M_{\rm BH}\sim9\times10^{8}M_{\odot}$ \citep{3c84_mass}.
The combination of proximity and large black hole mass makes 3C\,84 an ideal candidate to 
study the jet base and its polarization properties with the highest possible angular and spatial resolution, using global mm-VLBI.
For the adopted distance an angular scale of 50$~\mu$as corresponds to a linear scale of $\sim0.0175~$pc or $\sim203$ Schwarzschild radii ($R_{s}$).

At 1.3\,mm and 0.9\,mm wavelengths (230 and 341\,GHz), \cite{plambeck14} report 
a typical degree of linear polarization $m_L\gtrsim1$\,\%
using the Combined Array for Research in Millimeter Astronomy (CARMA) 
and the Submillimeter Array (SMA) with arc-second scale angular resolution. 
The authors also find a large rotation measure of RM$\sim9\times10^{5} \rm rad/m^{2}$. 
IRAM 30m Telescope observations at 86 and 230~GHz reveal a typical 
RM~$\sim7\times10^{4} \rm rad/m^{2}$ \citep{agudo14,agudo18b}. 
Both \cite{plambeck14} and \cite{agudo18b} report polarization variability on monthly timescales at mm-wavelengths, 
implying a small emission region (apparent size $\lesssim 1~c\times{\rm 1~month}\sim$0.025~pc $\sim 0.1$\,mas).

At cm-wavelengths and on milli-arcsecond scales, however, only low-level polarization is observed. 
As mentioned above, Very Long Baseline Array (VLBA) observations 
of the VLBI core of 3C\,84 reveal a linear polarization of $m_{L}\lesssim0.1$\% at 15\,GHz (upper limit) and $m_{L}\sim0.2$\% at 22\,GHz \citep{taylor06}. 
It was suggested that Faraday depolarization with
RM $\gtrsim10^{4} \rm rad/m^{2}$ could be an explanation \citep{taylor06}. The high RM is also consistent with
the non-VLBI measurements \citep{plambeck14,agudo18b}.
However, the limited angular resolution of the previous observations
does not yet allow to precisely locate the polarized emission region nor unambiguously determine the physical origin of the high RM.
Observations with the Global Millimeter-VLBI Array (GMVA) help to overcome this limitation. 
The observing frequency of 86\,GHz nicely bridges the gap between the 5-22\,GHz and 230\,GHz observations.
This will help to connect the results from the previous studies  
(e.g., \citealt{aller03,agudo10,trippe12,agudo18a,plambeck14,nagai17}), in order to form a
self-consistent physical jet polarization model from the cm- to the mm-regime.

In this paper, we present a new study of the polarization properties of 3C\,84
based on GMVA observations at 86\,GHz and supplemented by close-in-time VLBA observations at 43\,GHz and 15\,GHz.
In Sect. \ref{sec:method} we describe details of the VLBI observations and the data reduction.
Our main findings from the observations are summarized in Sect. \ref{sec:result}.
The physical implications are discussed in Sect. \ref{sec:discussion}.
In Sect. \ref{sec:conclusion} we summarize the results.

\begin{table*}[!t]
\caption{Summary of the polarimetric VLBI observations of 3C\,84 in May 2015.}              
\label{tab:summary}      
\centering                                      
\begin{tabular}{cccc ccc ccc}          
\hline\hline                        
Stations & Epoch & $\nu_{\rm obs}$ & Beam & $I_{\rm peak}$ & $S_{\rm tot}$ & $\sigma_{\rm I}$ & $PI_{\rm peak}$ & $P_{\rm tot}$ & $\sigma_{\rm P}$ \\    
(1) & (2) & (3) & (4) & (5) & (6) & (7) & (8) & (9) & (10) \\
   & [yy/mm/dd] & [GHz] & [mas, deg] & [Jy/bm] & [Jy] & [mJy/bm] & [mJy/bm] & [mJy] & [mJy/bm] \\
\hline                                   
VLBA(8)\\+GBT+EB+ON &  2015/05/16 & 86.252 &  0.048$\times$ 0.14 (-16.2) & 1.82 & 12.0 & 0.53 
& 27.7 &  42.0 &  5.03 \\
VLBA(10) 	   &   2015/05/11 & 43.115 & 0.15 $\times$ 0.30 (1.02)  & 3.82 & 17.0 & 1.24 &  12.3   & 16.8 & 1.75 \\
VLBA(10)           &   2015/05/18 & 15.352 & 0.42 $\times$ 0.63 (-8.74) & 4.00 & 28.8 & 1.30 & $<2.64$ & --   & 0.88 \\
\hline                                             
\end{tabular}
\tablefoot{
(1) Antennas participating in each epoch.
Numbers in the bracket indicates the number of the VLBA antennas participated.
(2) The observing epoch in year/month/day format.
(3) Center frequency.
(4) The restoring beam for the uniform weighting.
    Minor axis $\times$ major axis in mas (the position angle in degrees measured counterclockwise from North).
(5) Total intensity peak (in units of Jy per beam).
(6) Total flux density integrated over the entire VLBI image.
(7) Total intensity rms level.
(8) Linear polarization intensity peak
(upper limit for $3\sigma$ significance).
(9) Integrated linearly polarized flux density of the core region obtained by 
the quadratic sum of the integrated Stokes $Q$ and $U$ flux densities
in the region where significant linear polarization intensities are found
(``--'' means no significant linear polarization in the region displayed in Fig. \ref{fig:maps} and \ref{fig:gmva_inter_I}).
(10) The polarization intensity noise level obtained as described in Sect. \ref{subsub:pol_noise}.
}
\end{table*}

\section{Observations \& data reduction}\label{sec:method}

\subsection{GMVA 86\,GHz data}\label{subsec:gmva}

3C\,84 was observed with the GMVA at 86\,GHz in dual circular polarization in May 2015.
The source was observed as one of the $\gamma$-ray emitting AGN sources, which are monitored by semiannual GMVA observations 
that complement the VLBA-BU-BLAZAR monitoring program 
(\citealt{jorstad17}; for the GMVA monitoring see e.g., \citealt{rani15,hodgson17,casadio17}).
In this GMVA campaign a  total of 11 antennas observed 3C\,84:
Brewster (BR), Effelsberg (EB), Fort Davis (FD), the Green Bank Telescope (GB), Kitt Peak (KP), Los Alamos (LA),
Mauna Kea (MK), North Liberty (NL), Onsala (ON), Owens Valley (OV), and Pie Town (PT).
3C\,84 was observed in full-track mode for 8\,hrs. The data were recorded at an aggregate bitrate of 2\,Gbps with 2\,bit digitization.
Using the polyphase filterbank, the data were channelized in 8 intermediate frequencies (IFs) per polarization, with an
IF bandwidth of 32\,MHz, yielding a total recording bandwidth of 2$\times$256\,MHz.
The data were correlated using the DiFX correlator \citep{deller11}
at the Max-Planck-Institut f{\"u}r Radioastronomie (MPIfR) in Bonn, Germany.
A summary of details of the observations is given in Table \ref{tab:summary}.

\subsubsection{Total intensity calibration and imaging}\label{subsub:total_i}

The post-correlation analysis and calibration of the correlated data were done using standard VLBI fringe-fitting and 
calibration procedures in \textsc{AIPS} (\citealt{greisen90}; 
see also \citealt{martividal12} for the analysis of GMVA data).
In the first step of the phase calibration, delay and phase offsets between the IFs were removed
using high SNR fringe detection (so called manual phase-calibration).
After the phase alignment across the IFs, the global fringe fitting was performed over the full bandwidth,
in order to solve for the delays and rates combining all IFs.
We set the signal-to-noise (S/N) threshold to 5.
The number of detections was maximized when
we used a rather long fringe solution interval of 4 minutes, which was moved forward in time in 1 minute steps.
In the next step, we performed the a-priori amplitude calibration using the \textsc{AIPS} task \textsc{APCAL},
using a-priori calibration information provided by each station (system temperatures $T_{\rm sys}$ and 
elevation dependent gain curves). 
An opacity correction was performed for stations which did not apply it in the delivered system temperature values.
Finally, the cross-hand phase and delay offsets of the reference station were determined and removed using
\textsc{AIPS} task \textsc{RLDLY}. The UVFITS data were then exported outside 
AIPS using the \textsc{AIPS} task \textsc{SPLIT} by frequency-averaging the cross-power spectra.

At 86\,GHz, the atmosphere limits the phase coherence. In order to estimate the effect of coherence
losses, we fringe fitted the data once again using much shorter fringe solution intervals
comparable to the coherence time. 
For this purpose, we set the fringe solution interval to 10\,sec and adopted the same S/N threshold of 5.
We also used more narrow fringe search windows for delay and rate ($\pm10$\,ns and $\pm25$\,mHz, respectively).
The number of detected scans was reduced dramatically due to the short solution interval.
For the detected scans, we found that the short 10-second solution interval did not 
increase the baseline amplitudes significantly, as long as the phases were not averaged over a longer time interval.
(e.g., $\lesssim10$\% of decoherence amplitude loss with 10\,sec time-averaging).

The frequency-averaged (but not yet time-averaged) data were read into the DIFMAP package \citep{difmap} for imaging.
We flagged outlying visibilities which were caused by known or previously unrecognized antenna problems
(such as pointing and/or focus errors, and bad weather). After an initial phase self-calibration using a point-source model,
the visibility data were coherently time averaged over 10 seconds. We then imaged the source structure in Stokes $I$ by applying
the standard, iterative CLEAN and self-calibration steps (for both phase and amplitude). 
With careful CLEAN and self-calibrations, our CLEAN model was able to reach the
maximum correlated flux density on the shortest uv-spacings (e.g., $\lesssim200\,{\rm M}\lambda$).
We however note that the limited accuracy of the a-priori calibration for the stations which form the shortest baselines
(e.g., PT and LA) could have an effect on the maximum absolute flux density.

In order to correct for a systematic flux density scaling, we combined quasi-simultaneous data from VLBA observations of the source 
at 15 and 43\,GHz (Sect. \ref{subsec:low_vlbi}) and computed the synchrotron spectrum of the extended jet in the source 
(at $\sim2-4$\,mas from the VLBI core).
We then assumed that the observed spectral index of the extended jet components between 15 and 43\,GHz
is constant (no spectral break), and thus is the same as at the higher frequencies (i.e., between 43 and 86\,GHz).
Based on this assumption and the observed spectral index of $\alpha_{\rm 15-43\,GHz}\sim -0.5$ ($S\propto\nu^{+\alpha}$), 
we estimated a correction factor for the absolute flux density of the source and further calibrated the data.
We also checked the accuracy of the absolute flux density by making use of 
close-in-time single-dish observations and connected interferometer flux measurements of the source between 15 and 343.5\,GHz.
From this we conclude that the absolute flux density of the source in this epoch could be uncertain by $\sim30$\%
(see Appendix A for details).

We examined the reliability of the CLEAN model and the self-calibration solutions
by comparing the closure phases and amplitudes of the a-priori calibrated data and the CLEAN model.
In particular, we compared the differences (and ratios) of the closure phases (and closure amplitudes) between the data and the model.
The residual closure phases and the closure amplitudes ratios were computed for
the histograms of the entire data.
We found that the closure quantities of the a-priori calibrated data and the CLEAN model agreed
within $\sim10^{\circ}$ for the phases and $\sim25$\% for the amplitude, respectively.

\begin{figure}[t!]
\centering
\includegraphics[width=0.48\textwidth]{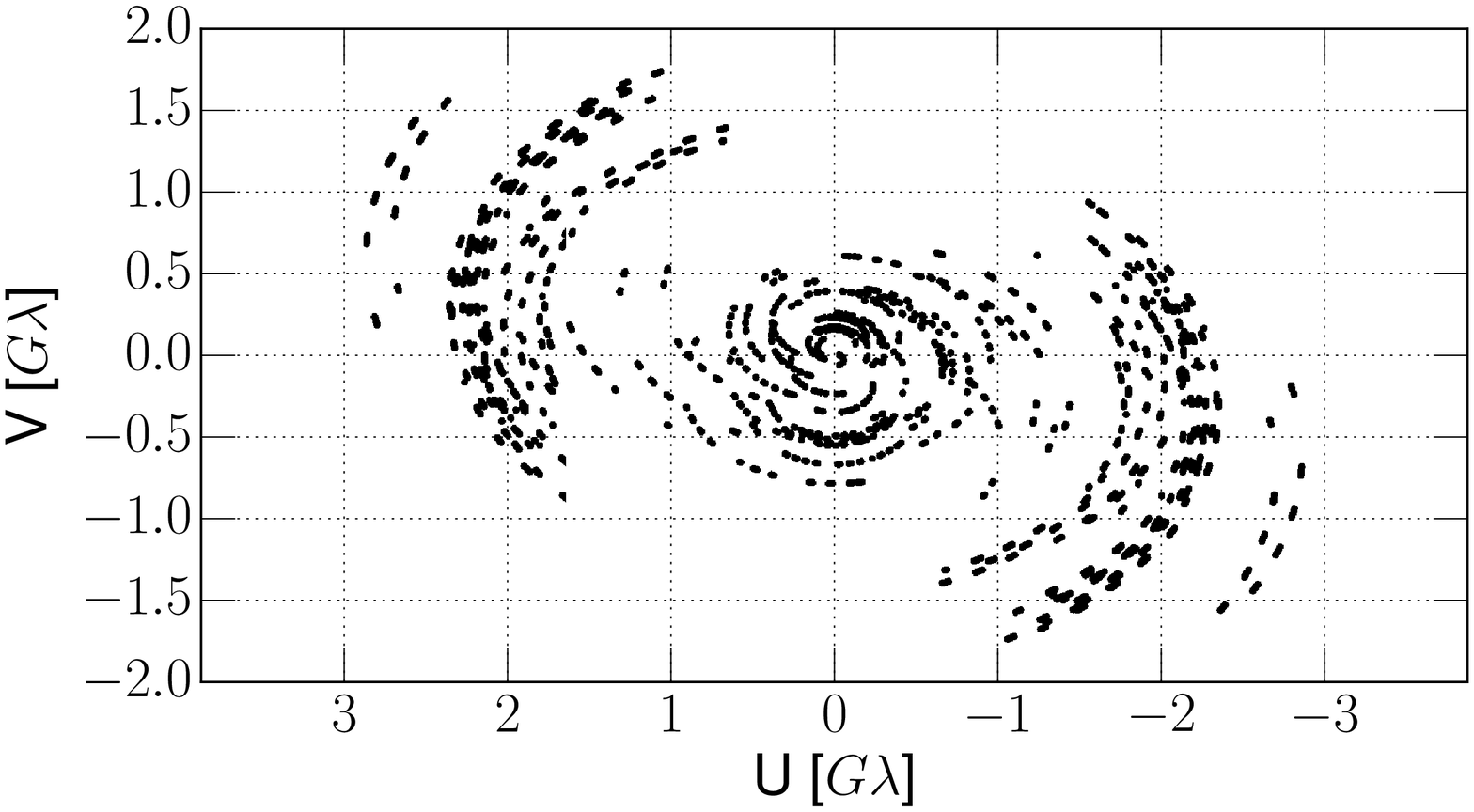}
\includegraphics[width=0.48\textwidth]{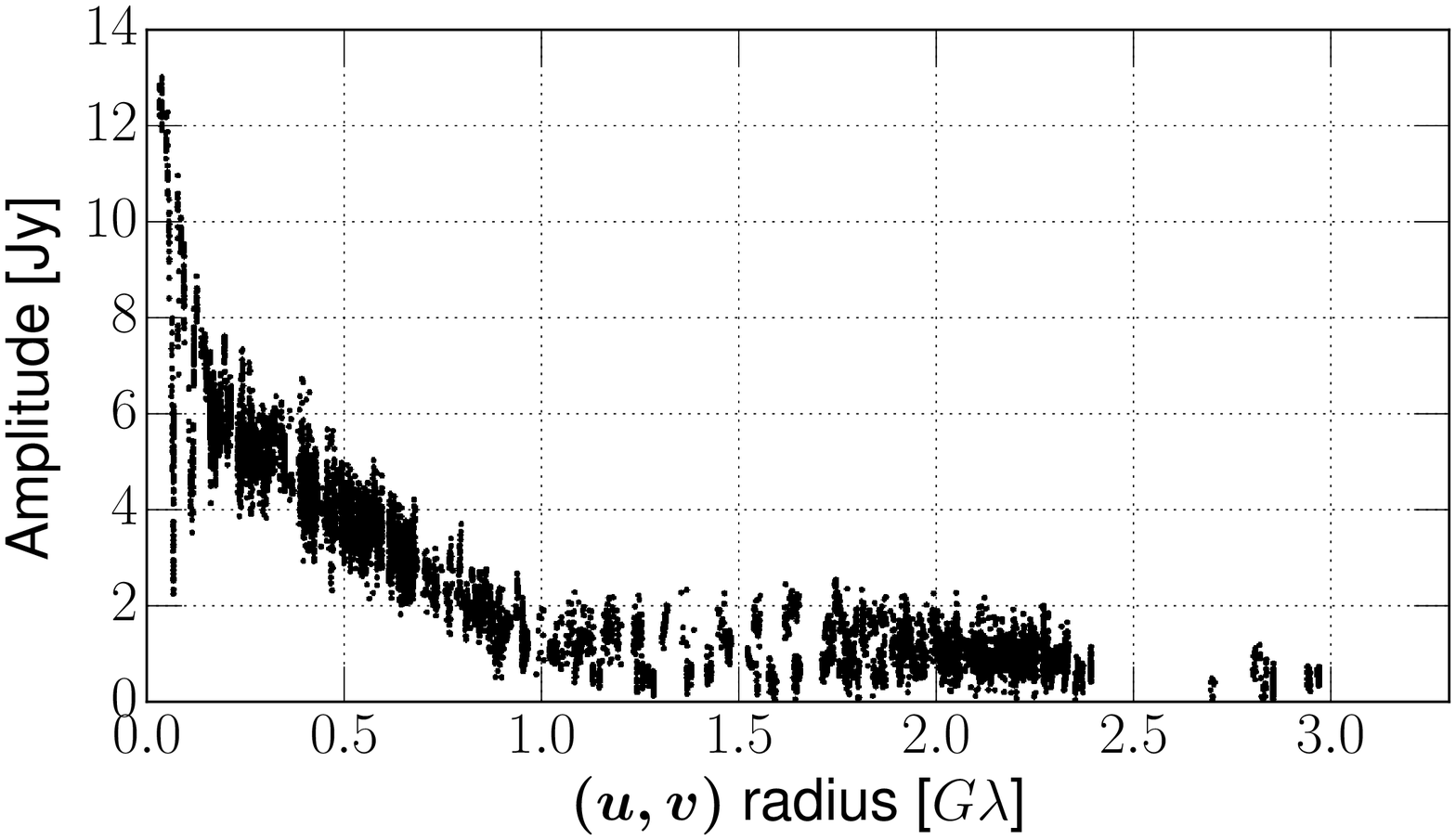}
\caption{
Stokes $I$ visibilities of the GMVA data.
{\it Top} : The $(u,v)$-coverage after the a-priori calibration and additional flagging for the imaging.
{\it Bottom} : Radial distribution of the visibility amplitudes after the last round of CLEAN and self-calibration.
The visibility data were binned in 30\,sec for clarity.
}
\label{fig:uv_rad}
\end{figure}

A summary of characteristics of the final 86\,GHz Stokes $I$ image is given in Table \ref{tab:summary}.
In Fig. \ref{fig:uv_rad} we show 
the $(u,v)$ coverage of the data set and the visibility amplitudes versus the $(u,v)$-distances after all the calibrations.

\subsubsection{Polarization calibration and imaging}\label{subsub:pol_cal}

After the total intensity imaging, additional calibration steps were performed in \textsc{AIPS} 
in order to remove residual gain offsets between the RCP and LCP receivers and the polarization leakage effects.
In particular, the antenna instrumental polarization solutions (D-terms) 
were determined using the \textsc{AIPS} task \textsc{LPCAL} \citep{leppanen95}.
We derived the D-terms from other more compact calibrator sources
including 3C\,454.3, BL\,Lac, CTA\,102, and OJ\,287, which were observed in the same session.
We also derived the D-terms from 3C\,84 itself, which is facilitated by (i) its  
core-dominace at millimeter wavelengths (see Sect. \ref{sec:result}) and (ii) 
wide parallactic angle coverage in this data set.
We show two linear polarization images of 3C\,84 in Appendix \ref{appendix:imaging_test} 
for D-terms solutions derived from CTA\,102 and OJ\,287. 
The maps show a broadly consistent appearance of the polarization. 
In order to obtain a final set of D-terms, we vector-averaged the D-terms 
and created the final polarization image of 3C\,84 for further analysis.
Confidence in the final D-terms was obtained by imaging the polarized structure of the calibrators.
Their GMVA polarization images showed good agreements with contemporaneous VLBA 43\,GHz polarization images
available from the VLBA-BU-BLAZAR database\footnote{https://www.bu.edu/blazars/VLBAproject.html}, 
but with more fine-scale structure seen in the GMVA images. Those images will be published later elsewhere.

The final D-term solutions are tabulated in Table \ref{tab:dterms}.
Their errors are estimated based on the standard deviation of the different D-term values for each station and polarization.

The absolute electric vector position angle (EVPA) of 3C\,84 was calibrated using OJ\,287, 3C\,454.3, and BL\,Lac. 
These calibrators were also observed as part of the POLAMI 
(Polarimetric Monitoring of AGN at Millimeter Wavelengths) program\footnote{http://polami.iaa.es/} \citep{agudo18a}
by the IRAM 30m Telescope, which was equipped with the XPOL polarimeter \citep{thum08}.
We estimate the overall uncertainty of the EVPA measurements in GMVA data using the following procedure:
first, the measurement error of the 86\,GHz EVPA in the IRAM 30m Telescope data is $\sim2-7$\,deg 
for different sources and for this particular epoch (see also \citealt{agudo18a}).
By deriving the VLBI EVPA correction values from multiple calibrators, we adopt a conservative absolute EVPA uncertainty of $\sim10$\,deg. 
Second, 
the polarization angle of the source and the D-term phases are correlated 
and uncertainties of the latter could affect the accuracy of the former (see, e.g.,  Eq. 3 and 13 of \citealt{leppanen95}).
The average phase errors of the D-terms is $\sim24$\,deg
(Table \ref{tab:dterms}).
Accordingly, we estimate that the corresponding EVPA uncertainty would be $\sim24/2=12$\,deg.
Third, 
we compute the peak-to-noise of the polarization features in Stokes Q and U and propagate the errors to estimate
the EVPA uncertainty due to the thermal noise.
Under the assumption that these three measurements errors are uncorrelated, 
we can add them in quadrature. For a signal-to-noise of $\sim5$,
this leads to the overall EVPA uncertainty of $\sim$17\,deg in this GMVA observation.

\subsubsection{Estimating noise levels in the polarization image}\label{subsub:pol_noise}

The Rice bias correction for polarization measurements (see, e.g., \citealt{wardle74,montier15_a,montier15_b})
requires an accurate estimation of the signal-to-noise ratio.
We simulated the probability density distribution of the noise in the linear polarization intensity image, $\sigma_{\rm P}$, 
by performing a Monte-Carlo simulation. First,
we computed the $1\sigma$ noise levels for Stokes $Q$ and $U$ by fitting Gaussian functions to pixel value histograms of 
source-free regions in the Stokes $Q$ and $U$ images.
We then simulated Gaussian-distributed noises for $Q$ and $U$ and added them in quadrature
to obtain the probability distribution of the polarization noise.
By calculating the 68\%, 97\%, and the 99.7\% confidence levels, we obtained the noise values of $5.7$, $9.5$, and $14.0$\,mJy/beam
for the conventional 1$\sigma$, 2$\sigma$, and 3$\sigma$ levels, respectively. 
Based on this, we adopt $1\sigma_{\rm P}\sim5\,$mJy/beam at 86\,GHz. 

We note that this value is higher than the noise in the Stokes $I$ image ($\sigma_{\rm I}\sim0.5$\,mJy/beam).
In the ideal case, the noise level of a linear polarization map, $\sigma_{\rm P}$, 
is expected to be comparable to or lower than that of the total intensity map, $\sigma_{\rm I}$,
because the latter is often  dynamic-range limited (see, e.g., \citealt{tms}).
However, this is not the case if systematic polarization calibration errors dominate the polarization noise
(e.g., due to errors in the D-terms).
Thus, we checked if the increased noise level in the linear polarization map could be due to uncertain D-term solutions.
Following \cite{roberts94} (see also \citealt{hovatta12}), we estimate the error from D-term calibration
as follows:

\begin{equation}
\sigma_{\rm Dterm}=
\frac{\Delta m}{  \sqrt{N_{\rm ant}\times N_{\rm IF} \times N_{\rm scan} } }
\sqrt{
I^{2} + (0.3\times I_{\rm peak})^{2}
}
\label{eq:dterm_noise}
\end{equation}

\noindent where 
$\Delta m$ is the error of the D-term amplitudes,
$N_{\rm ant}$ is the number of the antennas,
$N_{\rm IF}$ is the number of IFs,
$N_{\rm scan}$ is the total number of independent scans with different parallactic angles,
$I$ is the total intensity pixel values, and
$I_{\rm peak}$ is the peak value of the total intensity image.
As shown in Table \ref{tab:dterms}, 
the D-term amplitudes are uncertain by $\sim2-4$\% 
(the largest uncertainty of $\sim5$\% is found for RCP at PT). 
Therefore, we assume that a characteristic value of $\Delta m$ is $\sim$0.03.
For the GMVA data of 3C\,84, we adopt $N_{\rm ant}=11$ and $N_{\rm IF}=8$.
The number of independent scans including all the antennas was $N_{\rm scan}=8$.
This gives us a characteristic $\sigma_{\rm Dterm}\sim10^{-3}\times I_{\rm peak}$.
For $I_{\rm peak}=1.8$\,Jy/beam, we obtain $\sigma_{\rm Dterm}\sim2$\,mJy/beam.
The order of $\sigma_{\rm Dterm}$ appears to be in agreement with our estimation of $\sigma_{\rm P}$, suggesting that
residual polarization leakages may dominate the polarization noise in our 86\,GHz image.

\begin{table}[!t]
\caption{
Summary of the antenna D-terms at 86\,GHz.
The Green Bank Telescope was used as the reference antenna.
The columns show
(1) the station code,
(2,3) the amplitude ($m$) and phase ($\chi$) of the D-terms for the RCP,
and
(4,5) the same for the LCP.
}              
\label{tab:dterms}      
\centering                                      
\begin{tabular}{@{\extracolsep{-0.3pt}}ccccc@{}}          
\hline\hline                       
Station & \multicolumn{2}{c}{RCP} & \multicolumn{2}{c}{LCP} \\\cline{2-3}\cline{4-5}
     & $m$ & $\chi$ & $m$ & $\chi$ \\
  & [\%] & [deg] & [\%] & [deg] \\
 (1) & (2) & (3) & (4) & (5) \\
\hline                                   
BR & $5.8\pm2.7$ & $-(106\pm16)$  & $7.2\pm2.7$& $-(48\pm20)$  \\
EB & $4.5\pm2.1$ & $56\pm24$  & $2.3\pm2.5\tablefootmark{a}$ & $98\pm40\tablefootmark{a}$   \\
FD & $7.9\pm2.9$ & $23\pm13$  & $7.0\pm3.6$& $-(141\pm25)$  \\
GB & $1.7\pm2.4\tablefootmark{a}$ & $-(162\pm46)\tablefootmark{a}$  & $2.6\pm1.3$ & $-(66\pm44)\tablefootmark{a}$ \\
KP & $3.1\pm2.8$ & $178\pm30$  & $3.8\pm1.7$ & $88\pm14$ \\ 
LA & $10.7\pm1.6$& $139\pm9$& $10.1\pm2.1$& $32\pm13$ \\
MK & $3.9\pm3.1$ & $11\pm25$  & $4.3\pm1.9$ & $-(90\pm23)$ \\
NL & $4.8\pm1.6$ & $-(167\pm21)$   & $3.7\pm1.3$ & $70\pm13$  \\
ON & $5.9\pm1.9$ & $-(178\pm27)$  & $4.8\pm2.4$ & $-(4\pm36)$ \\
OV & $3.6\pm2.4$ & $-(48\pm38)$   & $5.4\pm1.7$ & $-(127\pm21)$  \\
PT & $8.2\pm5.6$ & $7\pm8$  & $9.6\pm3.4$ & $-(134\pm21)$  \\
\hline                                             
\end{tabular}
\tablefoot{
\tablefoottext{a}{Small D-term amplitudes induce large uncertainties in the D-term phases.}
}
\end{table}

\subsection{Contemporaneous 43 and 15\,GHz VLBA data}\label{subsec:low_vlbi}

On short timescales of  $\lesssim$\,one week, the structural variability of 3C\,84
is likely negligible because of the known slow apparent jet speed of $0.1c \sim 0.1$\,mas/yr $= 2 \mu$as/week in the core region of 3C\,84 \citep{walker94,suzuki12}.
This allows to combine the 86\,GHz data with quasi-contemporaneous observations at other frequencies.

We made use of archival polarimetric VLBA data of 3C\,84 provided by
the MOJAVE program\footnote{https://www.physics.purdue.edu/MOJAVE/} (15\,GHz; \citealt{lister05}) and
the VLBA-BU-BLAZAR monitoring program (43\,GHz; \citealt{jorstad17}).
We have chosen a set of 15\,GHz and 43\,GHz data which were obtained closest in time to the GMVA observations. 
The archival data were observed with the same dual polarization setup (RCP and LCP) and the same total bandwidth of 
$2\times256$~MHz.
The Stokes $I$, $Q$, and $U$ maps again were obtained using \textsc{DIFMAP}, averaging the visibilities over the whole observing bandwidth.
At these two frequencies, we estimate the polarization image noise levels $\sigma_{\rm P}$ in the same way as described in Sect. 
\ref{subsub:pol_noise}. 
The $\sigma_{\rm P}$ values were comparable to $\sigma_{\rm I}$,
indicating that at these lower frequencies the systematic uncertainties in the linear polarization due to errors in the D-terms are small.
A summary of the two data sets is given in Table \ref{tab:summary}.

For the absolute EVPA measurement error
(i.e., those arising from the D-term calibration and absolute EVPA correction),
we adopted $5$\,deg and $10$\,deg at 15\,GHz and 43\,GHz based on \cite{lister05} and \cite{jorstad05}, respectively.
The overall uncertainty for the EVPA was then estimated in the same way as we described in Sect. \ref{subsub:pol_cal}
by adding the systematic and thermal errors in quadrature.

In addition, the VLBA-BU-BLAZAR program provides the VLBA 43\,GHz data at four different frequencies in the 43\,GHz band
(43.0075, 43.0875, 43.1515, and 43.2155\,GHz). 
Their polarization calibration has been performed separately for each IF 
with respect to the D-terms and EVPA values (see Sect. 3.1 of \citealt{jorstad05}).
Therefore, we created additional 43\,GHz polarization images of the source for each of the four IFs separately,
in order to resolve the $n\pi$ angle ambiguity in the rotation measure analysis 
and to check for a possible large EVPA rotation inside the 43\,GHz band.

\begin{figure*}[t]
\centering
\includegraphics[height=0.55\textwidth]{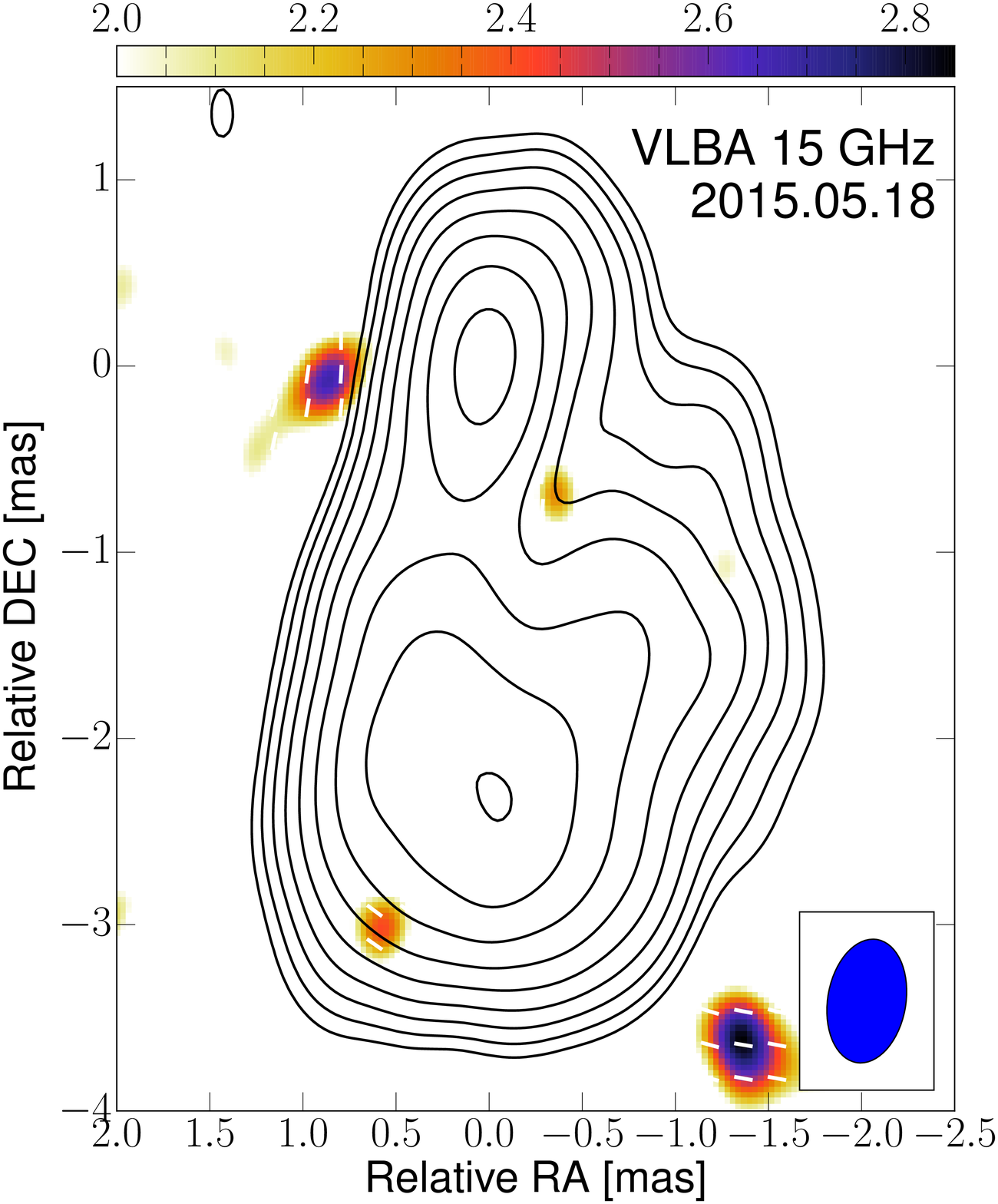}
\includegraphics[height=0.55\textwidth]{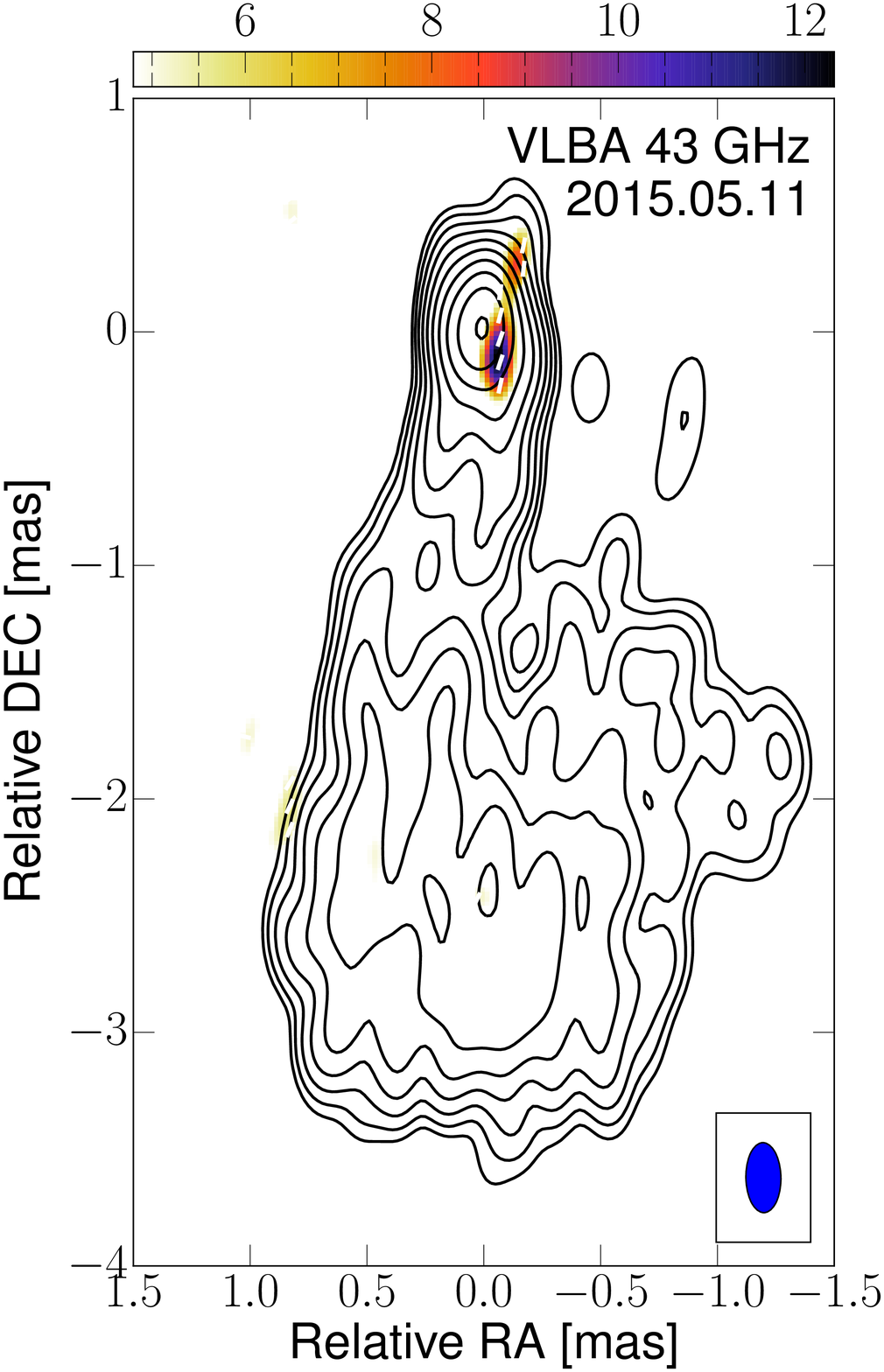}
\caption{
Polarimetric VLBI images of 3C\,84 at 15 and 43\,GHz.
In both panels, the contours show the total intensity and increase by a factor of 2 
from the lowest level.
The color scale represents the polarized intensity (in mJy/beam). 
The white bars in the images display the apparent EVPA.
The blue ellipse in the bottom right corner shows the restoring beam.
\textit{Left} : VLBA 15\,GHz image.
The contours start at 30\,mJy/beam.
The color scale starts at 2.0\,mJy/beam.
\textit{Right} : VLBA 43\,GHz image.
The contours start at 7\,mJy/beam.
The color scale starts at 4.8\,mJy/beam.
}
\label{fig:maps}
\end{figure*}

\subsection{Non-contemporaneous VLBA data at 43\,GHz}\label{subsec:other_43g}

We also analyzed other VLBA 43\,GHz data of 3C\,84 available from the BU database
in order to study the time variability of the total flux and the linear polarization over a longer period.
We specifically focused on observations performed in 2015 
(eight epochs; Feb, Apr, May, Jun, Jul, Aug, Sep, and Dec).
We imaged the source again by applying some additional data flagging when  
systematically large errors were clearly seen in some antennas or scans.
The full-polarization images available from the BU database reveal a weak polarization feature
in the core region of 3C\,84 in all the eight epochs during 2015. 
In our imaging, we recover a very similar polarization morphology in most epochs, however 
some epochs (Feb, Jun, Aug, and Dec) appear to be more limited in polarization fidelity,
possibly due to the limited number of scans or other systematic errors.
Therefore, we did not include these polarization images in our analysis.
Overall, repeatedly appearing and therefore reliable polarization features have been detected in four (Apr, May, Jul, and Sep) out of the eight epochs.
We also measured the unpolarized core flux for all the eight epochs 
using the MODELFIT routine implemented in DIFMAP
and analyzed their uncertainties following \cite{schinzel12} (see Sect. \ref{subsec:fit_err}). 
We summarize details for these data in Appendix \ref{appendix:43ghz}.

\subsection{Contemporaneous ALMA data}\label{subsec:alma}

Finally, we made use of contemporaneous (May 2015) linear polarization measurements of 3C\,84
from the Atacama Large Millimeter/submillimeter Array (ALMA) 
Calibrator Source Catalogue\footnote{
https://almascience.eso.org/alma-data/calibrator-catalogue
}.
The source was observed at 97.5, 233.0, and 343.5\,GHz 
and was not resolved up to the baseline length of $537.5\,k\lambda$ at all frequencies.
Therefore, we constrain the source size to be $<1/(537.5\times10^{3})~{\rm rad}\approx0.38$~arc second.
Table \ref{tab:alma} summarizes the archival ALMA polarization measurements.


\begin{table}[!t]
\caption{
Summary of the ALMA archival data.
The columns show
(1) the observing epoch,
(2) the central frequency,
(3) the total flux,
(4) the degree of the linear polarization, and
(5) the EVPA.
}              
\label{tab:alma}      
\centering                                      
\resizebox{0.5\textwidth}{!}{%
\begin{tabular}{ccccc}          
\hline\hline                        
Epoch & $\nu_{\rm obs}$ & $S_{\rm tot}$ & $m_{L}$ & EVPA \\
 (1) & (2) & (3) & (4) & (5) \\\relax
 [yyyy/mm/dd] & [GHz] & [Jy] & [\%] & [deg] \\
\hline                                   
2015/05/31 & 97.5 & $17.14\pm0.57$ & $0.6\pm0.3$ & $82.1\pm19.7$ \\
& 233.0 & $9.82\pm0.33$ & $1.0\pm0.3$ & $-(29.1\pm12.1)$ \\
& 343.5 & $6.85\pm0.23$ & $0.3\pm0.3$ & $18.3\pm42.9$ \\
\hline                                             
\end{tabular}
}
\end{table}

\subsection{Model-fitting, polarization measurements, and their uncertainties}\label{subsec:fit_err}

In order to parameterize the size and the flux density of the VLBI core region in Stokes $I$,
we made use of the MODELFIT procedure implemented in DIFMAP.
We fitted circular Gaussians to the visibilities at each frequency
and estimated their uncertainties following \cite{schinzel12}.
For the absolute flux density at 86\,GHz, we conservatively adopt  an uncertainty of $\sim30$\% 
based on the limitation of the absolute flux calibration in this epoch.

The polarization components were identified by the 
local maxima of their linearly polarized intensities in the image plane (see Fig. \ref{fig:gmva_inter_I} and \ref{fig:gmva_inner}). 
For each component, we define a circular aperture whose diameter matches the mean beam size
and is centered at the local peak of the polarized intensity.
The properties of the polarized components -- 
i.e., the Stokes $I$ flux density $S$, 
the linearly polarized flux density $P$, 
the degree of linear polarization $m_{L}$, and
the EVPA $\chi$ --
were then determined from the spatially integrated Stokes $I$, $Q$, and $U$ flux densities.
Specifically, the integrated flux densities and the polarization parameters were computed by

\begin{eqnarray} 
Q_{\rm tot} &=& \Sigma Q_{i,j} \times A_{\rm pixel}/A_{\rm Beam} \quad \textrm{(in Jy)}
\label{eq:measurement_q} \\
U_{\rm tot} &=& \Sigma U_{i,j} \times A_{\rm pixel}/A_{\rm Beam} \quad \textrm{(in Jy)} 
\label{eq:measurement_u} \\
S &=& \Sigma I_{i,j} \times A_{\rm pixel}/A_{\rm Beam} \quad \textrm{(in Jy)} 
\label{eq:measurement_s} \\
P &=& \sqrt{Q_{\rm tot}^{2}+U_{\rm tot}^{2}} \quad \textrm{(in Jy)} 
\label{eq:measurement_p} \\
m_{L} &=& 100\times P/S \quad \textrm{(in percent)} 
\label{eq:measurement_ml}\\
\chi &=& 0.5\arctan(U_{\rm tot}/Q_{\rm tot}) \quad \textrm{(in radian)}
\label{eq:measurement_chi}
\end{eqnarray}

\noindent where
$I_{i, j}$, $Q_{i, j}$ and $U_{i, j}$ are the corresponding Stokes intensity values at each pixel $(i, j)$ in Jy/beam
and
$A_{\rm pixel}$ and $A_{\rm Beam}$ are the pixel and the beam area in square mas, respectively\,\footnote{
We compute the area under the elliptical Gaussian beam, $A_{\rm Beam}$, by
$A_{\rm Beam}=\pi\psi_{\rm maj}\psi_{\rm min}/4\ln2$ where 
$\psi_{\rm maj, min}$ are the FWHM of the elliptical Gaussian along the major and the major axis,
respectively.
The areas of individual pixels, $A_{\rm pixel}$, are $0.01\times0.01$, $0.02\times0.02$, and $0.03\times0.03$\,square mas at 86, 43, and 15\,GHz, respectively.
}.
In these calculations, we also corrected for the Rice bias in the polarization intensity value at each pixel by following \cite{wardle74}.
We assume that the uncertainty of $P$ is $30$\% at all frequencies.
For $S$, we assume $10$\% uncertainties at $\leq43$\,GHz (e.g., \citealt{lister05,jorstad05}) 
but $30$\% at 86\,GHz as explained before.
The uncertainty of $m_{L}$ was then obtained by propagating the errors of the integrated flux densities.
The EVPA error was calculated as described in Sect. \ref{subsub:pol_cal}.

\section{Results}\label{sec:result}

\subsection{Total intensity structure and the core flux}\label{subsec:total_flux}

\begin{figure*}[t!]
\centering
\includegraphics[height=0.5\textwidth]{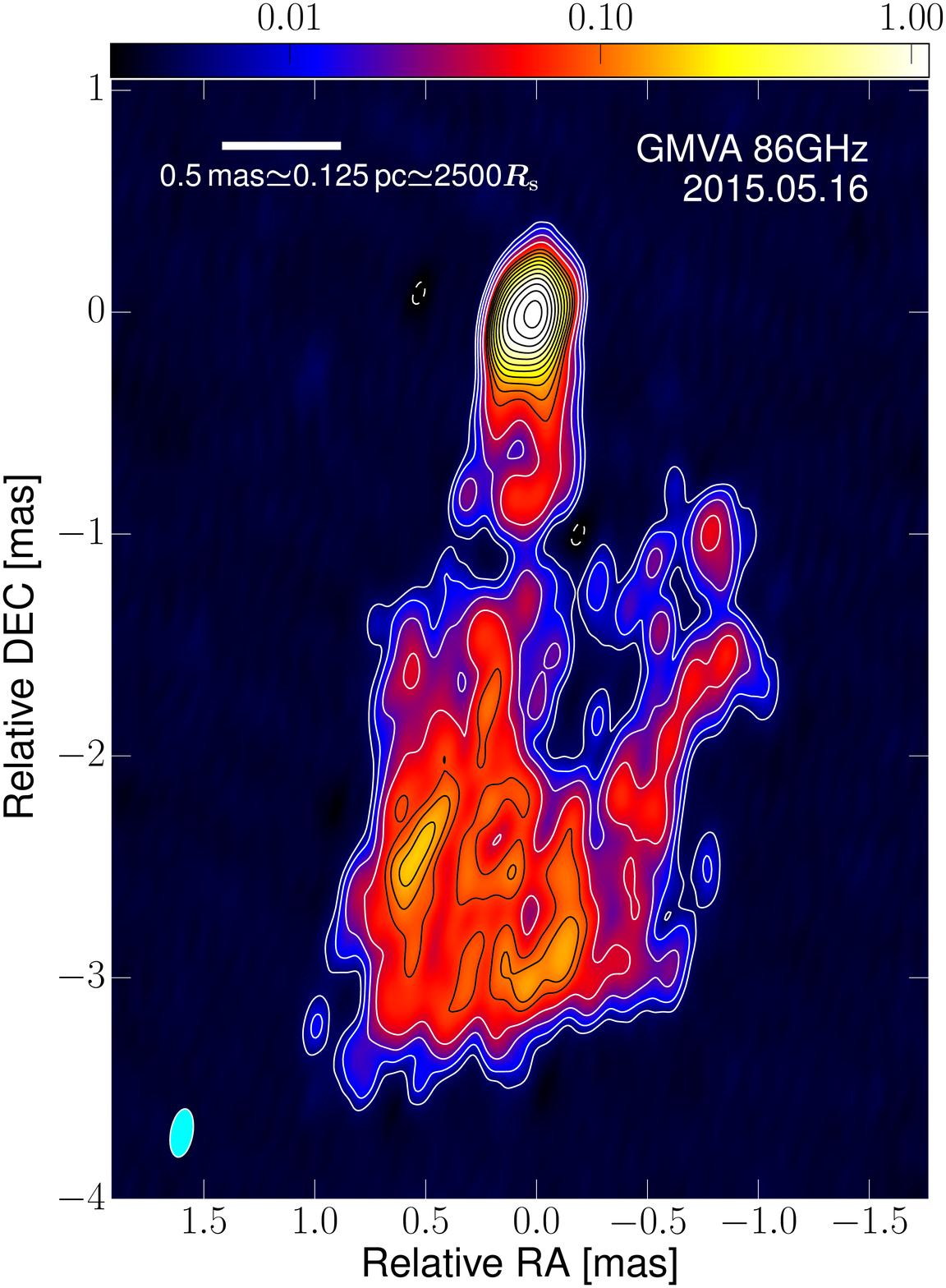}
\includegraphics[height=0.5\textwidth]{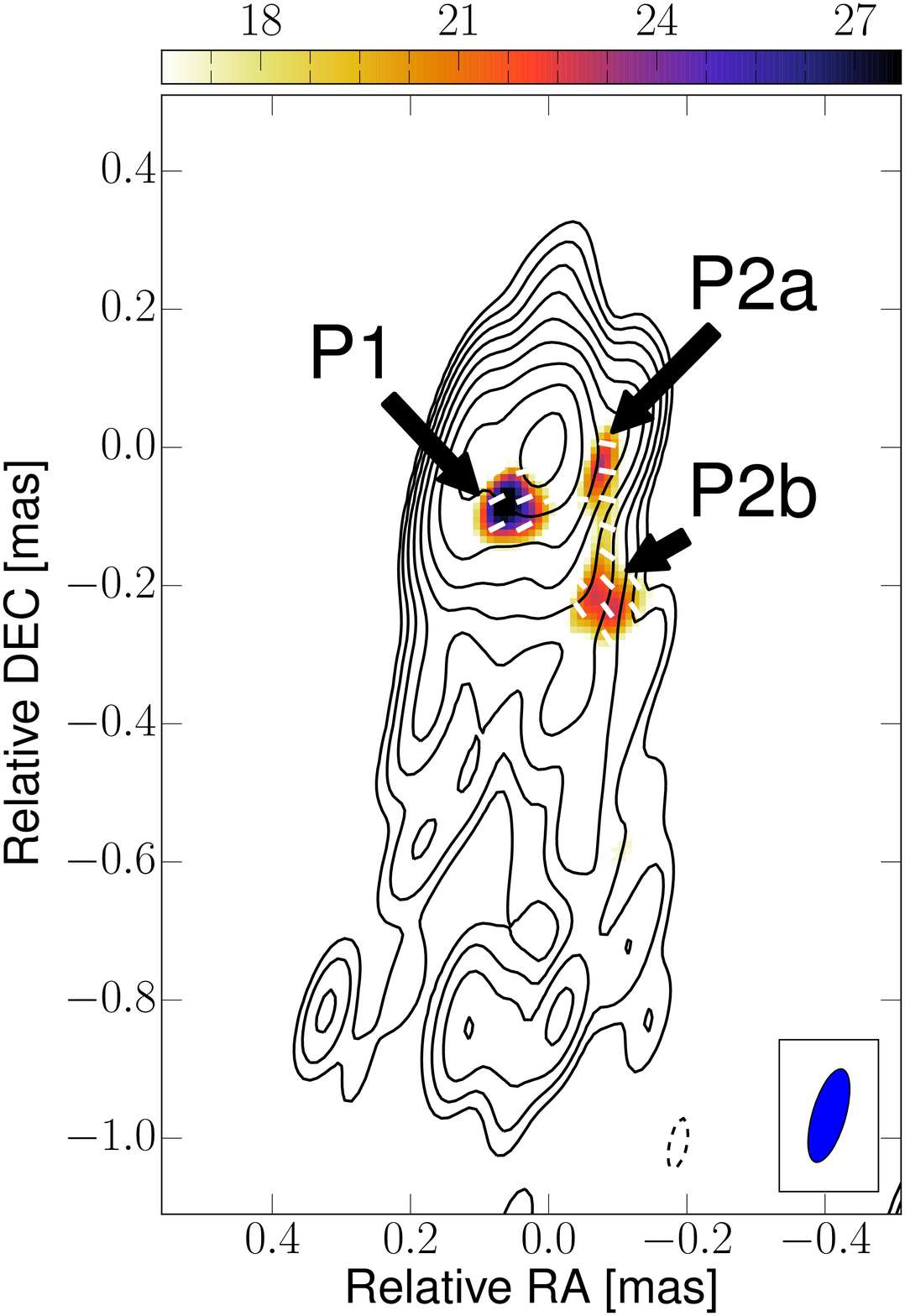}
\caption{
GMVA 86\,GHz images of 3C\,84.
In both panels, the contours show the total intensity and increase by a factor of 2 from the lowest level (5\,mJy/beam).
A negative contour of $-$5\,mJy/beam is also shown with dashed lines.
\textit{Left} :
The Stokes $I$ image restored with an elliptical beam of 
$0.22\times0.10$\,mas and position angle of $-10^{\circ}$.
This beam is larger than the full resolution attainable with the GMVA.
The peak of this image is 3.04\,Jy/beam. 
The cyan ellipse in the left bottom corner illustrates the elliptical beam.
The color scale represents the Stokes $I$ pixel values in Jy/beam.
The white bar and the text in the upper left corner denote the projected spatial scale of the image.
\textit{Right} : 
The inner jet structure obtained with the full angular resolution of the GMVA.
The color scale represents the linear polarization intensity and starts from 16.5\,mJy/beam.
       }
\label{fig:gmva_inter_I}
\end{figure*}

\begin{figure}[t!]
\centering
\includegraphics[width=0.45\textwidth]{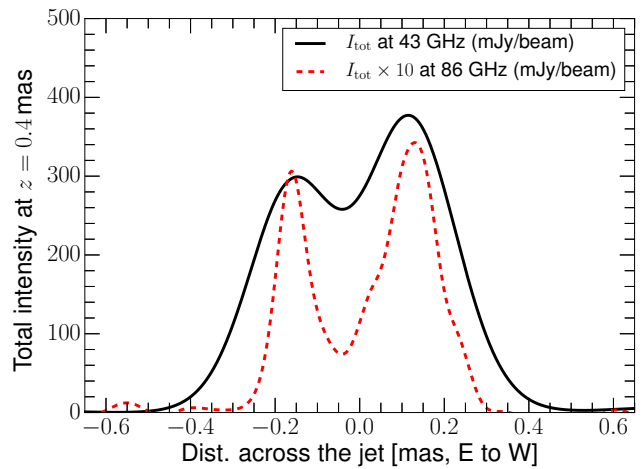}
\caption{
Transverse jet intensity profiles at 43\,GHz (black solid line) and 86\,GHz (red broken line) 
obtained at 0.4\,mas core distance
as function of distance across the jet (increasing towards west).
}
\label{fig:tot_slice}
\end{figure}

In Fig. \ref{fig:maps} we show the quasi-simultaneous polarimetric VLBI images of 3C\,84 at 15\,GHz and 43\,GHz in May 2015.
The GMVA 86\,GHz images are shown in Fig. \ref{fig:gmva_inter_I} with two different restoring beams
in order to better illustrate both the extended and inner jet structure.
The source clearly shows a limb-brightened jet at high frequencies ($\ge 43$~GHz).
The limb-brightening is visible up to $\sim2$~\,mas from the core.
We made transverse cuts to the jet at a core separation $z\sim0.4$\,mas at 43\,GHz and 86\,GHz 
using the full-resolution image (Fig. \ref{fig:gmva_inter_I}, right) and show the intensity profiles of the slices in Fig. \ref{fig:tot_slice}.
The limb-brightened transverse structure is consistent with previous deep single-epoch \citep{nagai14} and 
long-term \citep{jorstad17} VLBA 43\,GHz observations of the source. 
At 86\,GHz, the improved angular resolution resolves the edge-brightened structure down to $\sim0.2$\,mas core distance.
Remarkably, within the inner 0.2~mas from the intensity peak, the nuclear region appears substantially resolved 
in the E-W direction and shows a complex morphology.
The 86\,GHz core is well represented by a single circular Gaussian component 
whose FWHM size is comparable to the 43\,GHz core.
However, the visibilities at long baselines ($\sim1.0-2.5G\lambda$) 
suggest the presence of a more complex sub-nuclear structure (see Fig. \ref{fig:uv_rad}).
On larger scales, the overall source structure is quite similar to that shown in the 43\,GHz image
(cf. Fig \ref{fig:maps} and \ref{fig:gmva_inter_I}),
but the extended jet emission at larger core separations is overall fainter.
A more detailed study of the total intensity structure is beyond the focus of this paper and will be presented elsewhere.

\begin{figure}[t!]
\centering
\includegraphics[width=0.45\textwidth]{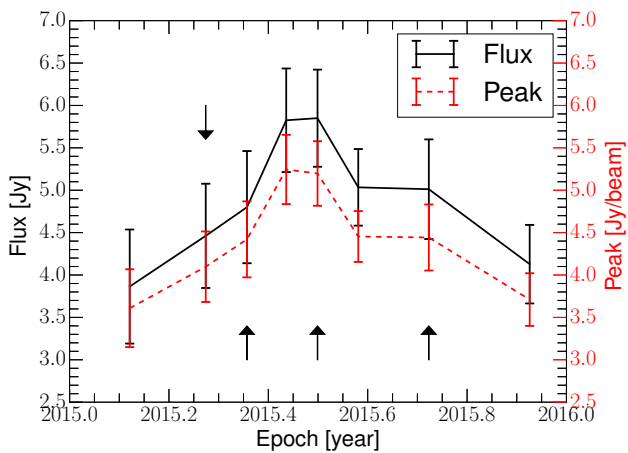}
\caption{
VLBA 43\,GHz core flux and the peak values during 2015.
Arrows indicate epochs when reliable polarization has been detected.
}
\label{fig:lc_43g}
\end{figure}

In Table \ref{tab:core} we show the flux densities $S_{\rm mod}$ and FWHM sizes of the core at 15, 43 and 86\,GHz
obtained with the MODELFIT procedure of DIFMAP.
We fitted a single power-law to the core spectrum, i.e. $S\propto\nu^{+\alpha}$ 
where $\nu$ is the observing frequency and $\alpha$ is the spectral index.
The results are shown in the left panel of Fig. \ref{fig:core_sed}.
We find that the spectrum of the VLBI core is slightly inverted up to 86\,GHz with a 
spectral index of $\alpha=0.51\pm0.10$.
We also show the time variation of the core flux at 43\,GHz in Fig. \ref{fig:lc_43g}.
In the same figure we also provide the peak values obtained with a circular Gaussian beam of $0.3$\,mas.
The core brightens by $\sim2$\,Jy during 2015 and reaches a local maximum during June and July 2015.


\subsection{Linear polarization in the core region}

\begin{figure*}[t!]
\centering
\includegraphics[width=0.44\textwidth]{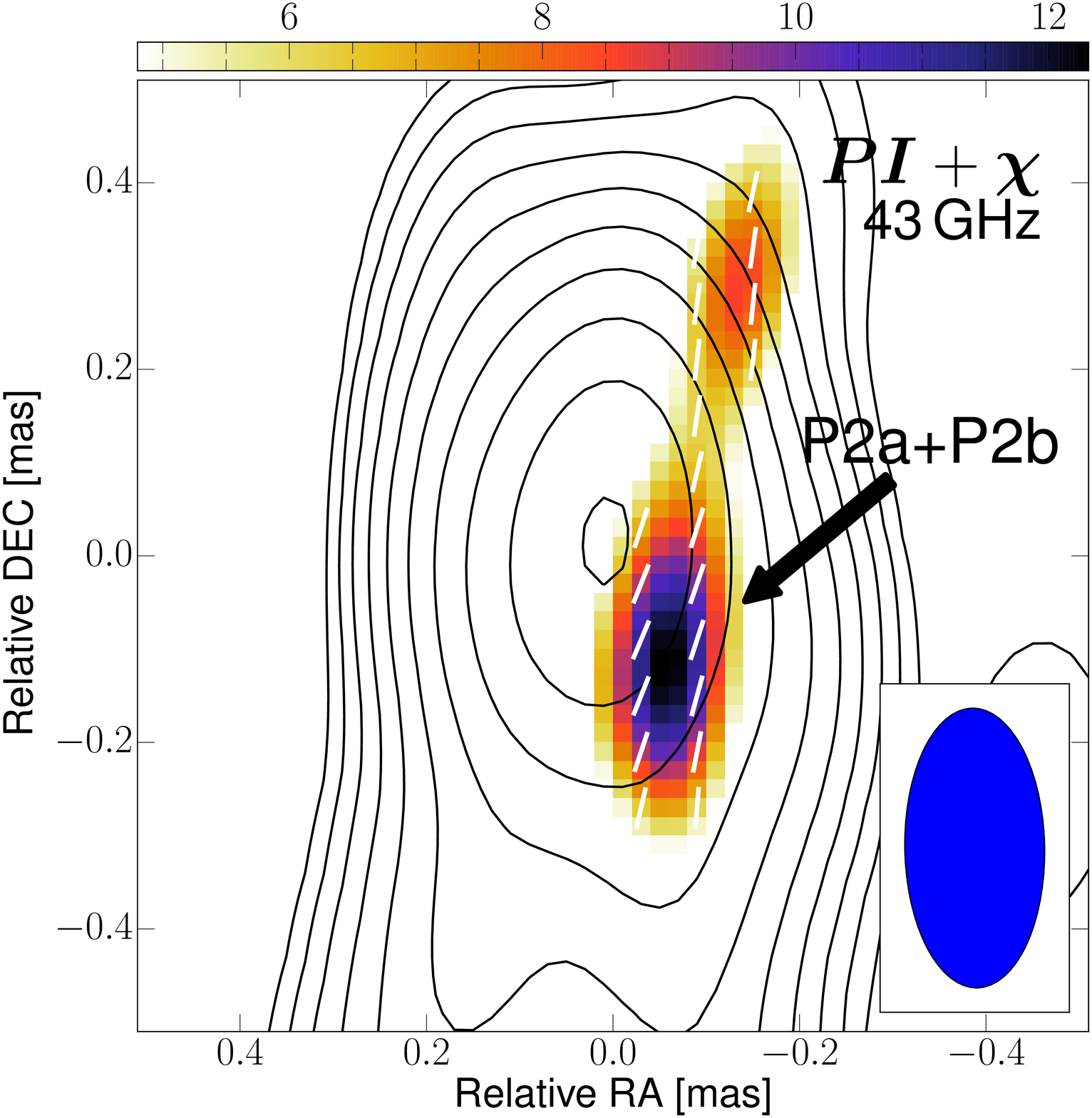}
\put(-190,190){\Large\bf (A)}
\includegraphics[width=0.44\textwidth]{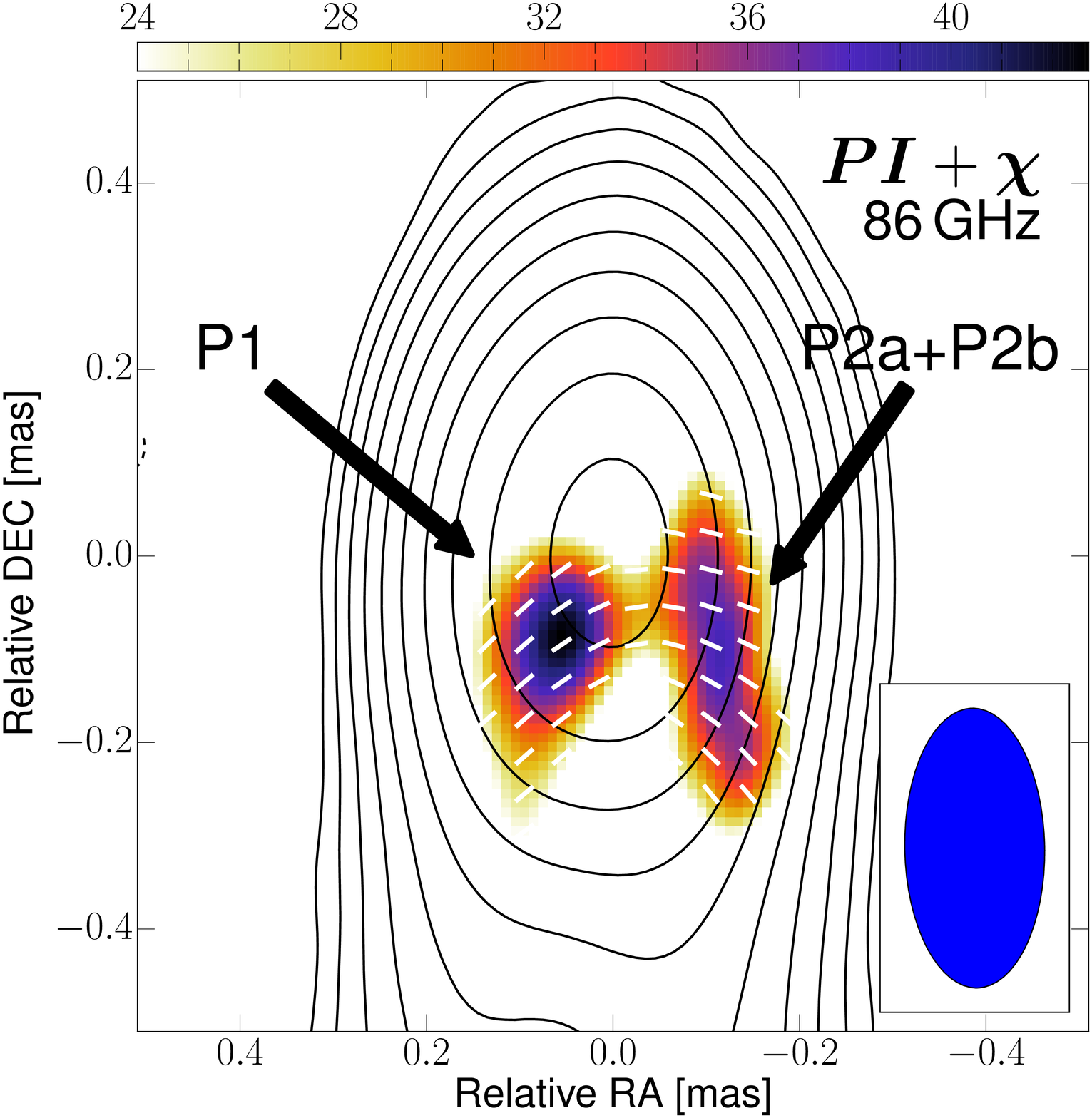}
\put(-190,190){\Large\bf (B)}\\
\caption{
Comparison of the total intensity, the polarized intensity ($PI$), and the polarization angle ($\chi$)
in the nuclear region at 43 and 86\,GHz.
The maps are restored with the same VLBA 43\,GHz CLEAN beam
and displayed over the same RA and DEC ranges ($\pm0.5$\,mas).
The arrows show the polarized component identifications.
(A) : The same as the central panel of Fig. \ref{fig:maps} (43\,GHz) but zoomed in the core.
(B) : GMVA 86\,GHz total and linear polarization image restored with the VLBA 43\,GHz beam.
The contours start at 6\,mJy/beam and increase by a factor of 2 from the lowest level.
The color scale starts from 24\,mJy/beam.
}
\label{fig:gmva_inner}
\end{figure*}

In Fig. \ref{fig:gmva_inner} we show
the linear polarization structure in the core region at 43 and 86\,GHz in more detail, 
restoring the maps with the VLBA 43\,GHz beam at both frequencies for better comparison.
In May 2015, we detect linear polarization features in the core 
at high significance ($PI_{\rm peak}/\sigma_{\rm P}\sim$ 7.0 and 5.5
at 43 and 86\,GHz, respectively; see Table \ref{tab:summary}).
At 15\,GHz, no significant polarization features appear in the nuclear region.
The polarization features at both 43 and 86\,GHz are slightly offset from the peak of the total intensity.
The features P1 and P2 (i.e., P2a+P2b) are separated by $\sim0.2$\,mas in the east-west orientation, and 
could possibly be associated with the edges of the underlying limb-brightened jet seen at 22\,GHz 
by recent RadioAstron observations (e.g., \citealt{giovannini18}). 
The P2 component is  present in both maps at similar locations.
However, the P1 component in the 86\,GHz image has no clear counter-part at 43\,GHz.
Using the pixel values at the position of P1, we estimate a $3\sigma_{\rm P}$ upper limit of $m_{L}$ 
of P1 is $\lesssim0.2$\% at 43\,GHz.
At 43\,GHz, there exists another polarized feature ($\sim8$\,mJy/beam) at a separation of $\sim0.3$\,mas 
to the north of the core, perhaps associated with a counter-jet.
We searched the VLBA-BU-BLAZAR database for evidence in support of significant polarization north to the main intensity peak
in several epochs.
While the polarization detection close to the total intensity peak can be seen in many epochs, 
we found polarization north of it only in a few epochs. 
Therefore we do not find strong evidence for a persistent polarization on the counter-jet side. 
Except for this faint northern polarization feature, the 43\,GHz VLBA polarization images in the other epochs 
show similar linear polarization flux densities in the core region, 
although the relative position of the polarization peak seem to change with time (see Appendix \ref{appendix:43ghz}).

At 86\,GHz, the degrees of the linear polarization $m_{L}$ of the polarized components are in the range of 
$\sim3-6$\% when convolved with the small GMVA observing beam.
The 86\,GHz $m_{L}$ values decrease to $\sim1-2$\% when convolved with the VLBA 43\,GHz beam.
For the P2 component, the $m_{L}$ value is significantly lower at 43\,GHz ($\sim0.4-0.5$\% in this epoch).
At 15\,GHz, the non-detection suggests a 3$\sigma$ upper limit to $m_{L}$ of $\lesssim0.1$\%,
which is in agreement with the results of \cite{taylor06}.
In order to minimize the effect of the different angular resolutions of
the VLBA and the GMVA images,
we now determine the polarization properties convolving the maps with the same VLBA 43\,GHz beam (Fig. \ref{fig:gmva_inner})
and compare the results in the following analysis.
Table \ref{tab:core_pol} provides
a summary of the properties of the polarized components in May 2015
(see Appendix \ref{appendix:43ghz} for results at 43\,GHz at other epochs).

\begin{figure*}[t!]
\centering
\includegraphics[width=0.45\textwidth]{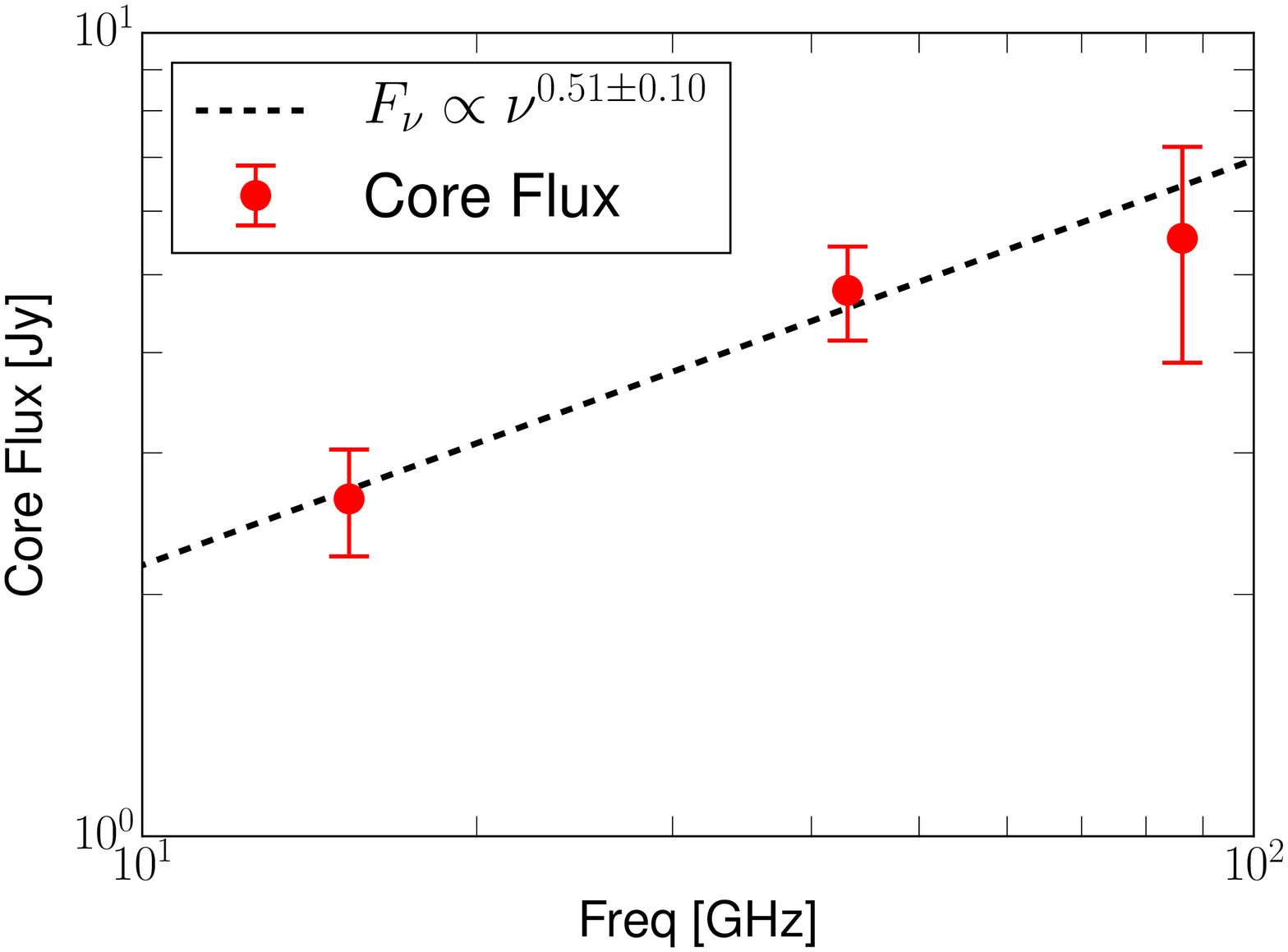}
\includegraphics[width=0.45\textwidth]{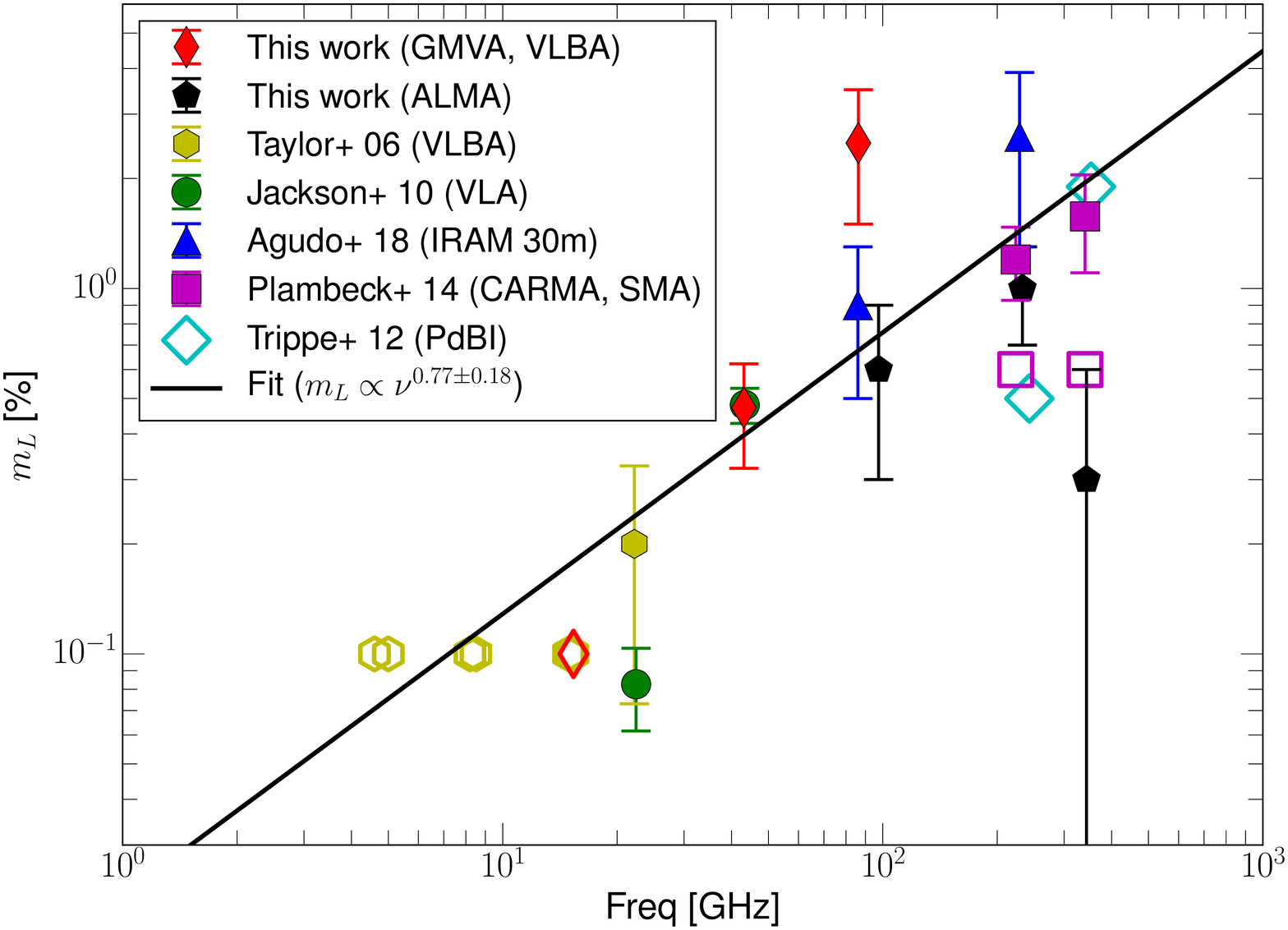}
\caption{
\textit{Left  }: Total intensity spectrum of the VLBI core region between 15 and 86\,GHz.
\textit{Right }: The degree of linear polarization in the VLBI core region obtained from 
this work and previous VLBA observations \citep{taylor06}
and integrated values from the literature \citep{jackson10,trippe12,plambeck14,agudo18b}.
All the VLBI polarization measurements were made from the pixel values near the peak of the Stokes $I$.
Filled and open symbols of the same marker indicate measured values and upper limits, respectively.
Citations and corresponding instruments are shown in the legend.
We note that at 230\,GHz only upper limits were observed before May 2011 \citep{trippe12,plambeck14}.
After October 2011, \cite{plambeck14} report a significantly higher linear polarization degree.
}
\label{fig:core_sed}
\end{figure*}

In order to investigate the frequency dependence of the linear polarization of the core region,
we combined linear polarization measurements from the literature with the data from this paper. 
The results are shown in the right panel of Fig. \ref{fig:core_sed}.
The linear polarization of the core clearly shows an increasing trend with frequency, which
can be represented by a power-law fit 
$m_{L} \propto \nu^{0.77\pm0.18}$, in which we have combined all available $m_{L}$ measurements.
We note that the index of the power-law could be affected by some systematic uncertainties
due to the different angular resolutions of the telescopes.
Nevertheless, we can clearly see that both VLBI and non-VLBI measurements show an 
increase of linear polarization degree towards higher frequencies.

\begin{table}[!t]
\caption{
VLBI core properties in May 2015.
The columns show
(1) the central observing frequency and
(2,3) the total flux and the FWHM size of the VLBI component
obtained by the Gaussian model-fitting.
}              
\label{tab:core}      
\centering                                      
\begin{tabular}{ccc}          
\hline\hline                        
$\nu_{\rm obs}$ &  $S_{\rm mod}$ & FWHM \\
 (1) & (2) & (3)  \\\relax
 [GHz] & [Jy] & [mas] \\
\hline                                   
86    &   $5.6\pm1.7$ & $0.13\pm0.04$ \\
43    &   $4.8\pm0.7$ & $0.14\pm0.01$ \\
15    &   $2.6\pm0.4$ & $0.27\pm0.02$   \\
\hline                                             
\end{tabular}
\end{table}

\subsection{Rotation measure between 43\,GHz and 344\,GHz in May 2015}\label{subsec:calc_rm}

We determine the rotation measure (RM) associated with the Faraday rotation
by measuring the variation of the observed EVPA with frequency.
When a linearly polarized wave at a wavelength $\lambda$ passes through magnetized plasma,
the intrinsic EVPA of the emission $\chi_{\rm int}$ is rotated and
the observed EVPA $\chi(\lambda)$ is

\begin{equation}
\chi(\lambda) = \chi_{\rm int} + {\rm RM} \times \lambda^{2} \quad
\label{eq:basic}
\end{equation}

\noindent
The rotation measure RM is determined by the gradient of 
$\chi(\lambda)$ versus $\lambda^{2}$ after solving for the $n\pi$ ambiguity in the angle.
To resolve the $n\pi$ ambiguity, 
we assume that most of the polarized flux observed by the ALMA at 97.5, 233.0, and 343.5\,GHz originates 
from the same VLBI core region.
This assumption appears to be reasonable based on the the flat spectrum of the 
linearly polarized flux density between 97.5 and 233.0\,GHz measured by ALMA
($\sim100$\,mJy at both frequencies; see Table \ref{tab:alma}).
Next, we compute the integrated Stokes $Q$ and $U$ flux densities of the VLBI core region 
at 43 and 86\,GHz to obtain the spatially integrated EVPAs.
The spatially integrated EVPAs within the 43\,GHz band are shown in Table \ref{tab:core_pol}.
At 86\,GHz, the integrated EVPA is $\chi=-(70\pm16)$\,deg. 
Then we compare all these EVPAs simultaneously to find a best $\lambda^{2}$ fit.

\begin{table}[!t]
\caption{
Properties of the polarized components in May 2015.
The columns show 
(1) the central frequency,
(2) the component identification,
(3) the Stokes $I$ flux density,
(4) the degree of linear polarization, and
(5) the EVPA.
The Stokes $I$ flux densities are consistent at the four separate frequencies
within the 43\,GHz band and we omit the numbers for the other three 43 GHz sub-bands.
}              
\label{tab:core_pol}      
\centering                                      
\resizebox{0.5\textwidth}{!}{%
\begin{tabular}{ccccc}          
\hline\hline                        
$\nu_{\rm obs}$ &  ID & $S_{\nu}$ & $m_{\nu} $ & $\chi_{\nu}$ \\
 (1) & (2) & (3) & (4) & (5) \\\relax
 [GHz] &   & [Jy] & [\%]  & [deg] \\
\hline                                   
86.252    & P1 &  $0.70\pm0.21$  &  $3.4\pm1.4$ & $-(54\pm17)$  \\
          & P1\tablefootmark{a}  &  $1.5\pm0.4$ & $1.4\pm0.6$ & $-(43\pm17)$ \\
          & P2a &  $0.27\pm0.08$  & $6.3\pm2.7$ & $-(89\pm17)$  \\
          & P2b &  $0.05\pm0.02$  & $44\pm20$\tablefootmark{b} & $47\pm17$\tablefootmark{b}   \\
          & P2a+P2b\tablefootmark{a} & $1.3\pm0.4$  & $2.2\pm0.9$ & $71\pm17$  \\
\hline
43.0075 & P2a+P2b &  $2.0\pm0.2$ & $0.44\pm0.14$ & $-(9\pm10)$  \\
43.0875 &               &                    & $0.49\pm0.16$ & $-(20\pm10)$ \\
43.1515 & 	        & 		     & $0.53\pm0.17$ & $-(24\pm10)$ \\
43.2155 &  		& 		     & $0.43\pm0.13$ & $-(15\pm10)$ \\
\hline
15.352  &        &      N/A & $<0.1$\tablefootmark{c} & N/A \\
\hline                                             
\end{tabular}
}
\tablefoot{
\tablefoottext{a}{Obtained from the image in panel (B) of Fig. \ref{fig:gmva_inner}.}
\tablefoottext{b}{The component is associated with the outer edge of the jet and could have higher 
systematic uncertainties than our estimation.}
\tablefoottext{c}{$3\sigma$ upper limit from the non-detection.}
}
\end{table}

We resolve the $n\pi$ ambiguity by following \cite{hovatta12}.
The authors determined the smallest possible EVPA rotations at each frequencies to achieve
a statistically acceptable fit with the $\lambda^{2}$ model.
Similarly, we rotated the 86.0, 97.5, 233.0, and 343.5\,GHz EVPAs by $(n_{1}, n_{2}, n_{3}, n_{4})\times\pi$ respectively
where $n_{i}\in(-30,-29,...,29,30)$ and $i\in(1,2,3,4)$. 
Then we calculated corresponding $\chi^{2}$ values for each rotation.
We note that $n=30$ corresponds to $\rm RM\sim10^{8}~{\rm rad/m^{2}}$ for the frequency range considered here.
$\rm RM\sim10^{8}~{\rm rad/m^{2}}$ is also equivalent to a single $180^{\circ}$ wrap within the 43\,GHz band.
Therefore, we are assuming that there is no RM $\gtrsim10^{8}~{\rm rad/m^{2}}$ in the source (cf. \citealt{plambeck14}).
For the same reason, we did not apply the $n\pi$ rotation within the 43\,GHz band.
Then we chose a set of $n_{i}$ which provided minimum $\chi^{2}$ value.

\begin{figure}[t!]
\centering
\includegraphics[width=0.5\textwidth]{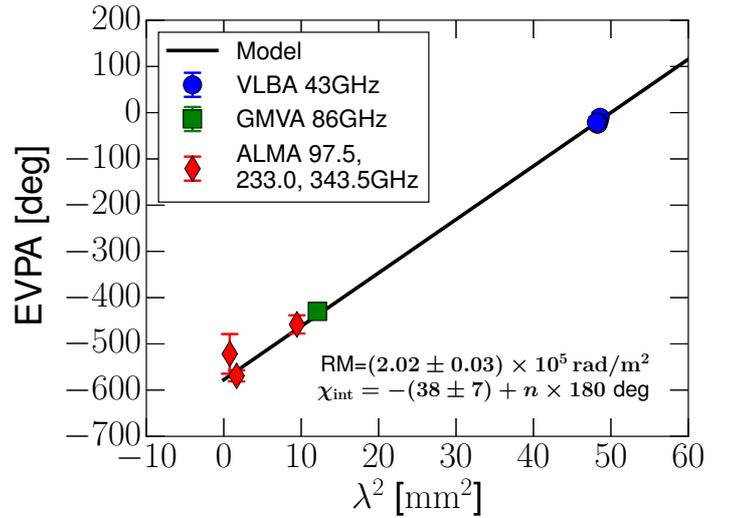}
\caption{
Spatially integrated EVPA versus the squared wavelength $\lambda^{2}$ (error bars) 
and the best $\lambda^{2}$ fit (solid line) in May 2015.
}
\label{fig:evpa}
\end{figure}

We find a best $\lambda^{2}$ fit with $n_{1}=-2$, $n_{2}=-3$, $n_{3}=-3$, and $n_{4}=-3$ 
(corresponding $\chi^{2}=3.52$ and the reduced $\chi^{2}=0.59$; see Fig. \ref{fig:evpa}).
Accordingly, we obtain ${\rm RM}=(2.02\pm0.03)\times10^{5}~{\rm rad/m^{2}}$ and 
$\chi_{\rm int}=-(578\pm7)$\,deg
(or equivalently $\chi_{\rm int}=-(38\pm7)$\,deg).
We note this RM is comparable to RM$\sim9\times10^{5}~{\rm rad/m^{2}}$ previously reported by \cite{plambeck14}.

\begin{figure}[!ht]
\centering
\includegraphics[width=0.45\textwidth]{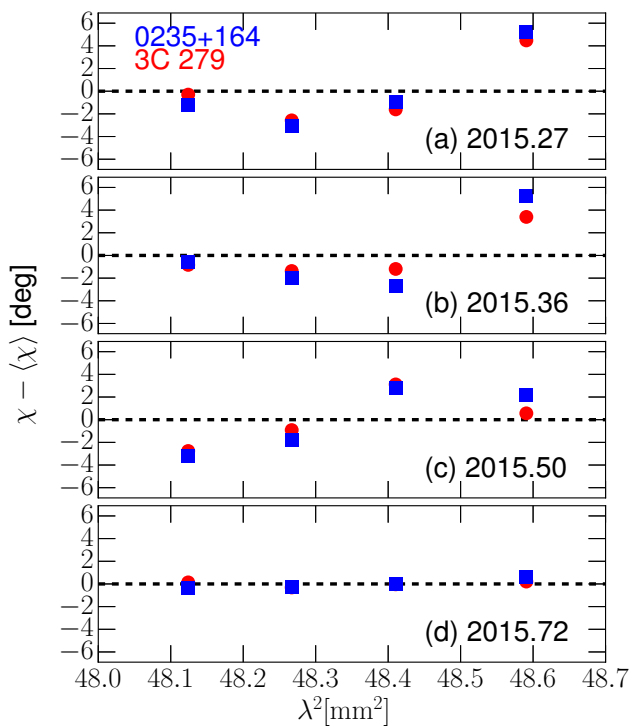}
\caption{
Variation of the EVPA in each sub-band, $\chi$, with respect to the mean EVPA, $\langle\chi\rangle$. 
Blue squares and red circles are for 0235+164 and 3C\,279, respectively.
Error bars are omitted for clarity. Note the strong correlation between the two sources.
}
\label{fig:evpa_diff_two}
\end{figure}

\subsection{Rotation measure within the 43\,GHz band}\label{subsec:rm_within_43}

We applied the same RM fitting procedure to the multi-epoch VLBA 43\,GHz data
using the EVPAs measured at the four different IFs.
We tested the significance of the EVPA rotation within the 43\,GHz band for 3C\,84 by performing
the same analysis to the core regions of two comparison sources 0235+164 and 3C\,279, 
which were observed in the same epochs by the VLBA-BU-BLAZAR monitoring program.
The instrumental polarization calibration and the sub-band EVPA calibration for the two sources 
were completely identical to those of 3C\,84 during 2015.
Therefore, we expect that the EVPA change within the 43\,GHz band 
will be systematically the same for the three sources 
if the EVPA changes are dominated by residual calibration errors.

\begin{table}[!t]
\caption{
RM within the VLBA 43\,GHz band.
From left to right, columns denote 
the observing epoch, the RM, and the EVPA interpolated to the central frequency of the 43\,GHz band (43.1155\,GHz).
}              
\label{tab:summary_rm_7mm}      
\centering                                      
\begin{tabular}{ccc}          
\hline\hline                        
Epoch & RM & $\chi$ \\\relax 
 [yyyy/mm/dd] & [$\times10^{5}~{\rm rad/m^{2}}$] & [deg] \\
\hline                                   
2015/04/12 & $-(6.2\pm2.6)$ & $11\pm5$ \\
2015/05/11 & $2.7\pm3.5$ & $-(14\pm5)$ \\
2015/07/02 & $6.8\pm3.3$ & $-(11\pm5)$ \\
2015/09/22 & $5.2\pm4.2$ & $-(8\pm5)$ \\
\hline                                             
\end{tabular}
\end{table}

In Fig. \ref{fig:evpa_diff_two} we show the EVPAs $\chi$ for 0235+164 and 3C\,279, with respect to their mean values, $\langle\chi\rangle$.
Within the 43\,GHz band, the EVPA of 3C\,279 varies only by $\leq6\,$deg in all epochs ($\leq8\,$deg for 0235+164).
This is insignificant compared to our presumed 10\,deg angle uncertainty at 43\,GHz. 
In addition, the two sources show strongly correlated EVPA offset from their mean values in all epochs.
From this we determine that the 43GHz in-band RM values smaller than $\sim(2-3)\times10^{5} \rm rad/m^{2}$ are insignificant
against the residual calibration errors.

\begin{figure}[t!]
\centering
\includegraphics[width=0.45\textwidth]{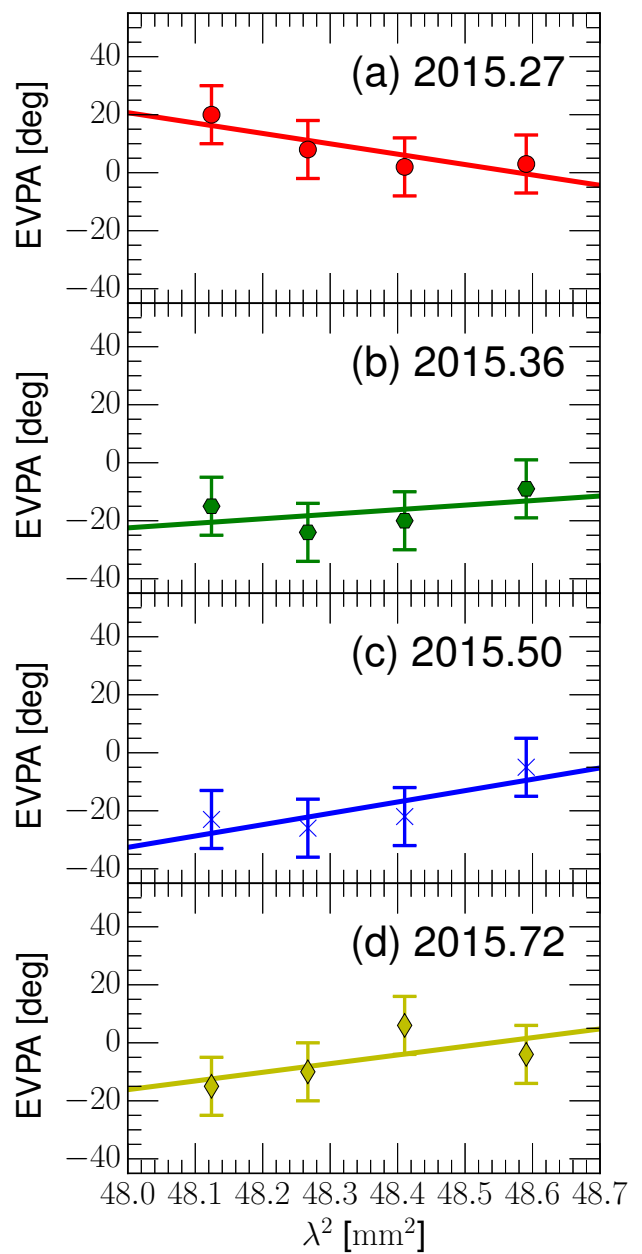}
\caption{
EVPA of 3C\,84 plotted versus the squared wavelength $\lambda^{2}$ within the VLBA 43\,GHz band (error bars) 
and the best-fit Faraday rotation model (solid line).
}
\label{fig:evpa_7mm_3c84}
\end{figure}

\begin{figure}[t!]
\centering
\includegraphics[width=0.5\textwidth]{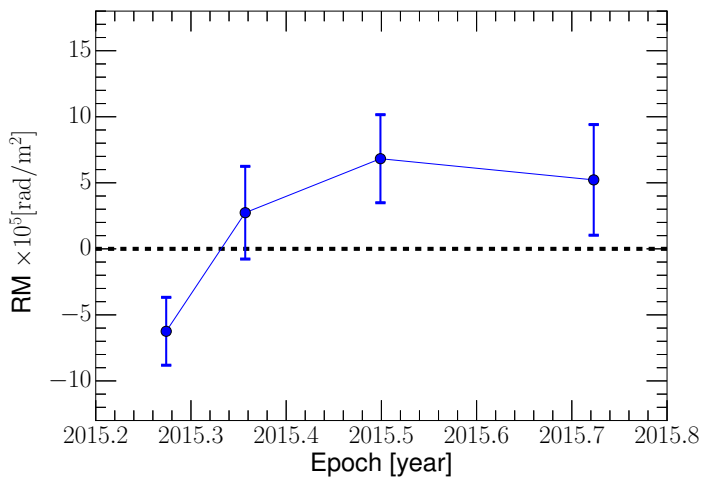}
\caption{
Time variation of the 43\,GHz sub-band RM for 3C\,84. 
}
\label{fig:7mm_rm_lightcurve}
\end{figure}

After this consideration, we show the EVPAs measured from the core of 3C\,84 at 43\,GHz band in Fig. \ref{fig:evpa_7mm_3c84}.
The corresponding RM values obtained from the fitting are tabulated in Table \ref{tab:summary_rm_7mm}
and a plot of the RM versus time is given in Fig. \ref{fig:7mm_rm_lightcurve}.
In contrast to 0235+164 and 3C\,279, 3C\,84 shows much larger EVPA variation within the 43\,GHz band ($15-21$\,deg).
These EVPA variation with wavelength is larger than the residual calibration errors that we estimated.
This results in a significant RM (${\rm |RM|}\gtrsim5\times10^{5}~{\rm rad/m^{2}}$ for three out of four epochs).
Accordingly, we consider the large RM obtained for 3C\,84 within the 43\,GHz band is intrinsic to the source.
We also note a discrepancy in the form of a negative RM 
(i.e., decreasing EVPA value with increasing $\lambda^{2}$), which is present only in Apr 2015.
We note that the EVPAs of the calibrators in this epoch generally increase with
increasing $\lambda^{2}$ (see the top panel of Fig. \ref{fig:evpa_diff_two}).
If the systematic EVPA rotation in the calibrators would be purely due to residual calibration errors, 
the intrinsic RM of 3C\,84 in this epoch would become even more negative.

\section{Discussion}\label{sec:discussion}

\subsection{Polarization structure}\label{subsec:morphology}

\begin{figure}[t!]
\centering
\includegraphics[width=0.45\textwidth]{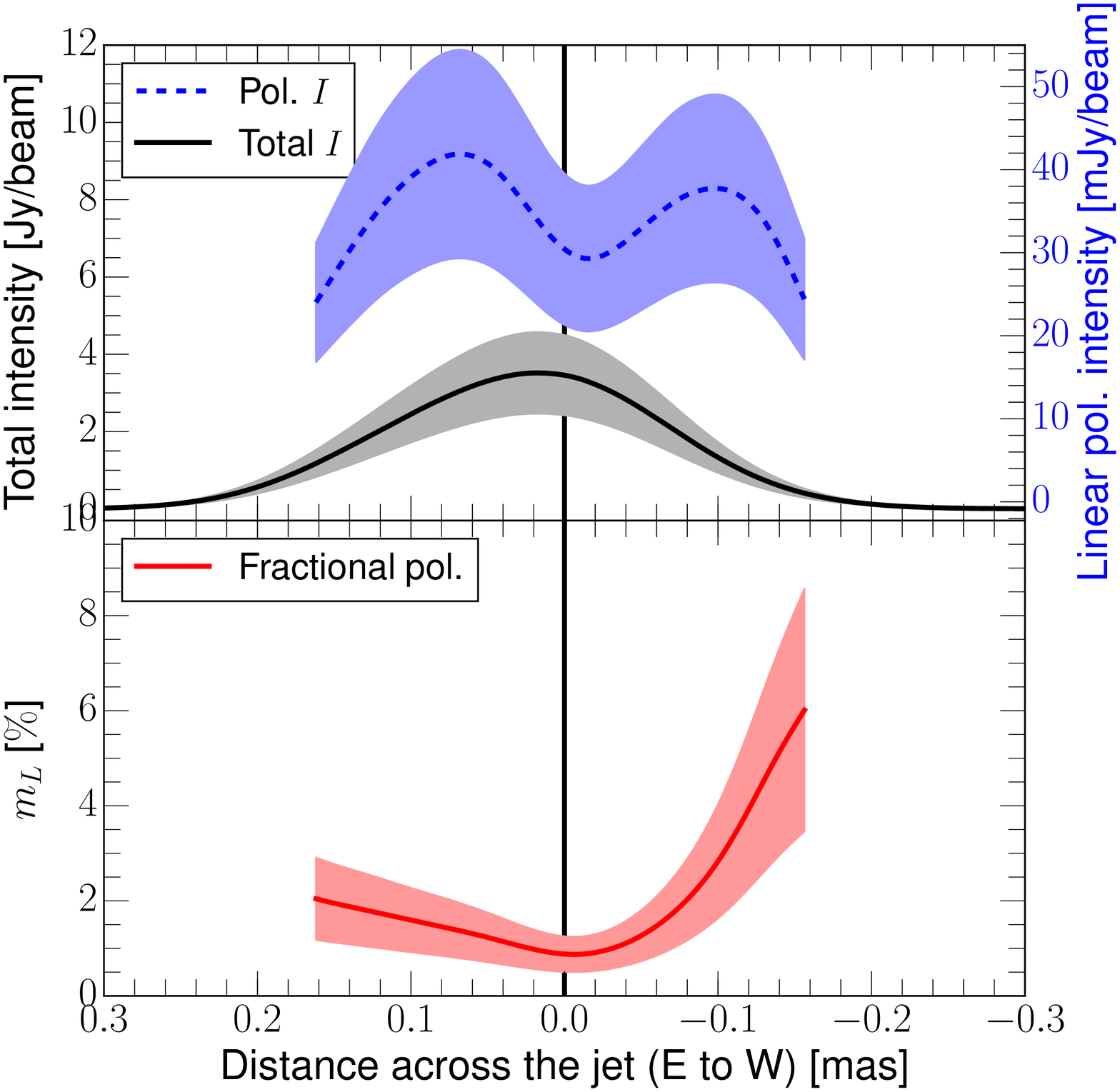}
\caption{
(Top) total intensity, linear polarization intensity and (bottom) degree of linear polarization at 86\,GHz
obtained by the slice transverse to the core and calculated by the pixel values in the image plane 
(negative core separation towards west; see Fig. \ref{fig:gmva_inner}).
The solid black line indicates position of the total intensity peak.
The shaded area marks the measurement errors, assuming 30\% uncertainty in the flux density measurements at 86~GHz 
(see Table \ref{tab:core}).
Only linear polarization higher than 24\,mJy/beam is shown.
}
\label{fig:slice}
\end{figure}

Here we briefly discuss possible implications of the polarization structure in the core region.
At 86\,GHz, the two sub-nuclear polarization features, 
P1 and P2, have different degrees of linear polarization. 
In order to highlight the polarization asymmetry, we chose the 86\,GHz image shown in Fig. \ref{fig:gmva_inner} and
made a transverse cut to the core region (in the E-W direction at $\sim0.1$\,mas downstream of the jet).
From this slice, we obtained transverse total and polarization intensity profiles (Fig. \ref{fig:slice}). 

A transverse polarization asymmetry in the core region has been occasionally seen in several other AGN  \citep{gomez08,osullivan09,clausen11,gomez16}.
This is often related to 
the geometry of the jet (e.g., jet inclination and opening angle) and/or
the magnetic field (e.g., pitch angle; see \citealt{lyutikov05,clausen11,porth11}).
We note that the intrinsic and spatially integrated EVPA is $\chi_{\rm int}\sim-38$\,deg in May 2015.
This angle is neither parallel nor perpendicular to the position angle of the inner jet ($\sim170$\,deg inferred from Fig. \ref{fig:gmva_inner}),
suggesting that the orientation of the magnetic field might be oblique to the jet. %

However, we also note that the polarization structure is also highly time-variable at 43\,GHz during 2015
when the VLBI core flux density showed a small outburst.
Such outbursts often accompany structural changes within the jet -- e.g., ejection of new VLBI components from the core --
and also likely lead to changes in the polarized morphology near the core. 
In fact, the peak of the linear polarization in the VLBA 43\,GHz images moves 
with respect to the peak of the total intensity by $\sim0.2$\,mas on a $\sim$monthly timescale,
especially in the E-W direction (see Appendix \ref{appendix:43ghz}).
We note that the typical VLBA beam at 43\,GHz beam is $\sim0.15$\,mas in E-W direction.
Thus the position offset of  $\sim0.2$\,mas is larger than the beam size and should be significant.
Single-dish monitoring observations of 3C\,84 at millimeter wavelengths by \cite{agudo18b} also show 
polarization time-variability on longer timescales.
Therefore, one cannot rule out the possibility of turbulence (e.g., \citealt{marscher14}) or varying opacity across the jet
which may also produce similar asymmetric and variable polarization structures.

\subsection{Possible explanations for the polarization spectrum and EVPA rotation}\label{subsec:pol_spec}

In the rest of this discussion, we focus on possible physical explanations for the observed frequency dependence of the 
fractional linear polarization and EVPA.
Frequency dependence of the polarization properties can be explained by several scenarios.
For instance, the inverted total intensity spectrum (Fig. \ref{fig:core_sed}, left) 
between 15 and 86~GHz suggests high opacity and large synchrotron self-absorption.
In such jets, the emission at a higher frequency originates from a region closer to the base of the jet (e.g., \citealt{lobanov98}).
If entanglement of the magnetic field would be the main source for the low linear polarization at low frequencies, 
a larger polarization degree $m_{L}$ at higher frequencies could be interpreted as progressively more ordered magnetic fields in the inner jet region
(e.g., \citealt{agudo14,agudo18b}). 
Alternatively, the large synchrotron opacity in the core itself can transform the observed polarization, 
making the linear polarization increasing with decreasing opacity (i.e., towards shorter wavelengths; see \citealt{pacholczyk,jones77}).
In particular, the transition from optically thick to optically thin is accompanied by a  $\sim90$\,deg flip of the EVPA 
\citep{pacholczyk}.

We point out, however, that the impact of Faraday rotation is also significant
because Faraday depolarization is a sensitive function of the observing frequency \citep{burn66,sokoloff98}.
Certainly, Faraday rotation may be not be a unique explanation of the large EVPA 
rotations across the observing frequencies.
However, we also find large EVPA rotation and the corresponding RM 
within the 43\,GHz band (Fig. \ref{fig:evpa_7mm_3c84}). 
This suggests that opacity effects alone cannot explain the results
because the opacity change within the 43\,GHz band is most likely small given the small 
fractional bandwidth of 256\,MHz/43\,GHz$\sim6\times10^{-3}$.

Within this perspective, the spectrum of the circular polarization in 3C\,84 is also noteworthy.
The inner jet of 3C\,84 has a large degree of circular polarization $m_{C}\sim1-3$\% at $15-22$\,GHz \citep{homan04}.
At higher frequencies, there are still no direct VLBI measurements of the circular polarization in this source.
However, single dish observations suggest that the fractional circular polarization has a flat spectrum between 
centimeter and millimeter wavelengths
($m_{C}\sim$ a few $\times0.1$\% at both $5-8$\,GHz and $86-230$\,GHz; \citealt{aller03,myserlis18,agudo10,thum18}).
As these authors suggest,
the combination of low fractional linear polarization and high circular polarization
is presumably produced by Faraday conversion,
which is accompanied by Faraday rotation in inhomogeneous plasma that is located within (or perhaps external) to the jet \citep{jones77,wardle03,macdonald16}.
Therefore, we investigate in the following the impact of Faraday rotation in terms of Faraday depolarization 
for different types of Faraday screens.

\subsection{Faraday depolarization}\label{subsec:depol}

Based on the previous discussion (Sect. \ref{subsec:pol_spec}),
we now discuss the impact of the Faraday effect assuming different types of Faraday screens.
For the central VLBI core region of 3C\,84, an external Faraday screen can be located at various places.
A free-free absorption disk \citep{walker2000}, which might be clumpy \citep{fujita17}, could act as a Faraday screen.
Another possibility is the accretion flow, which surrounds the central engine \citep{plambeck14}. We note that
a  radiatively inefficient accretion flow (RIAF) has been used to interpret the frequency dependence of the linear
and circular polarization of Sgr\,A* (e.g., \citealt{bower02,munoz12}), which resembles our observations of 3C\,84.
Finally, the host galaxy of 3C\,84 (NGC\,1275) contains a substantial amount of intergalactic gas, 
whose effect is seen on the pc-scale jet polarization at $\lesssim22$~GHz \citep{taylor06}.

In the following  we will model the Faraday depolarization for three very simple cases. We then will discuss their physical
implication in subsequent sections \ref{subsec:rm_external} and \ref{subsec:rm_internal}. Lets us assume three types of Faraday screens:
\begin{enumerate}
 \item A foreground screen with a disordered magnetic field and random RM fluctuations.
 \item A foreground screen with an ordered magnetic field and a smooth RM gradient within the observing beam.
 \item Faraday depolarization inside the jet and a uniform magnetic field.
\end{enumerate}
\noindent
Following \cite{burn66}, the observed degree of linear polarization $m_{obs}$ 
is a function of the observing wavelength, which for the three above cases can be written as:
\begin{eqnarray}
&{\rm Case~I : \quad } m_{obs} = & m_{0}\exp(-2\sigma^{2}\lambda^{4}) \label{eq:case1} \\
&{\rm Case~II : \quad } m_{obs} = & m_{0}\left|\frac{\sin(\Delta{\rm RM}\lambda^{2})}{\Delta{\rm RM}\lambda^{2}}\right| \label{eq:case2}\\
&{\rm Case~III : \quad } m_{obs} = & m_{0}\left|\frac{\sin(2{\rm RM}\lambda^{2})}{2{\rm RM}\lambda^{2}}\right| \label{eq:case3}
\end{eqnarray}
where 
$m_{0}$ is the intrinsic linear polarization degree,
$\lambda$ is the observing wavelength,
$\sigma$ is the standard deviation of the dispersion of the RM in the external Faraday screen,
$\Delta{\rm RM}$ is the RM gradient within the beam,
and RM is the observed rotation measure.
For Case I, we assume $\sigma\approx \rm \epsilon \times RM$ 
where $\epsilon$ is a scaling factor describing the level of the RM dispersion
(i.e., larger $\epsilon$ for more inhomogeneity).
We start the calculation with $\epsilon=1$.
Similarly, we assume $\Delta{\rm RM}\sim {\rm RM}$ for Case II.
For Cases II and III,
we have conservative upper limits for $m_{obs}$ by
$m_{obs} \leq m_{0}/(\Delta{\rm RM}\lambda^{2})$
and
$m_{obs} \leq m_{0}/(2{\rm RM}\lambda^{2})$, respectively.
We choose $m_{0}=10$\%, which is the theoretical maximum value in the case of high synchrotron opacity \citep{pacholczyk}.
Finally, we take a RM $\sim9\times10^{5}~{\rm rad/m^{2}}$ from \cite{plambeck14},
which is also close to the largest RM value found in our work ($\sim7\times10^{5}\rm rad/m^{2}$).
Then, we compare the upper limits with the observed $m_{L}$ values in May 2015.
For the P2 component, we use the $m_{L}$ values at 86 and 43\,GHz of Table \ref{tab:core_pol} 
and adopt $m_{L}\lesssim0.1$\% as an upper limit at 15\,GHz.
For the P1 component, we take the measured $m_{L}$ value at 86\,GHz and adopt $m_{L}\lesssim0.2$\% and $\lesssim0.1$\%
as upper limits for 43 and 15\,GHz, respectively.
For both components, we use the polarization degrees as measured with the VLBA 43\,GHz restoring beam.

\begin{figure}[t!]
\centering
\includegraphics[width=0.45\textwidth]{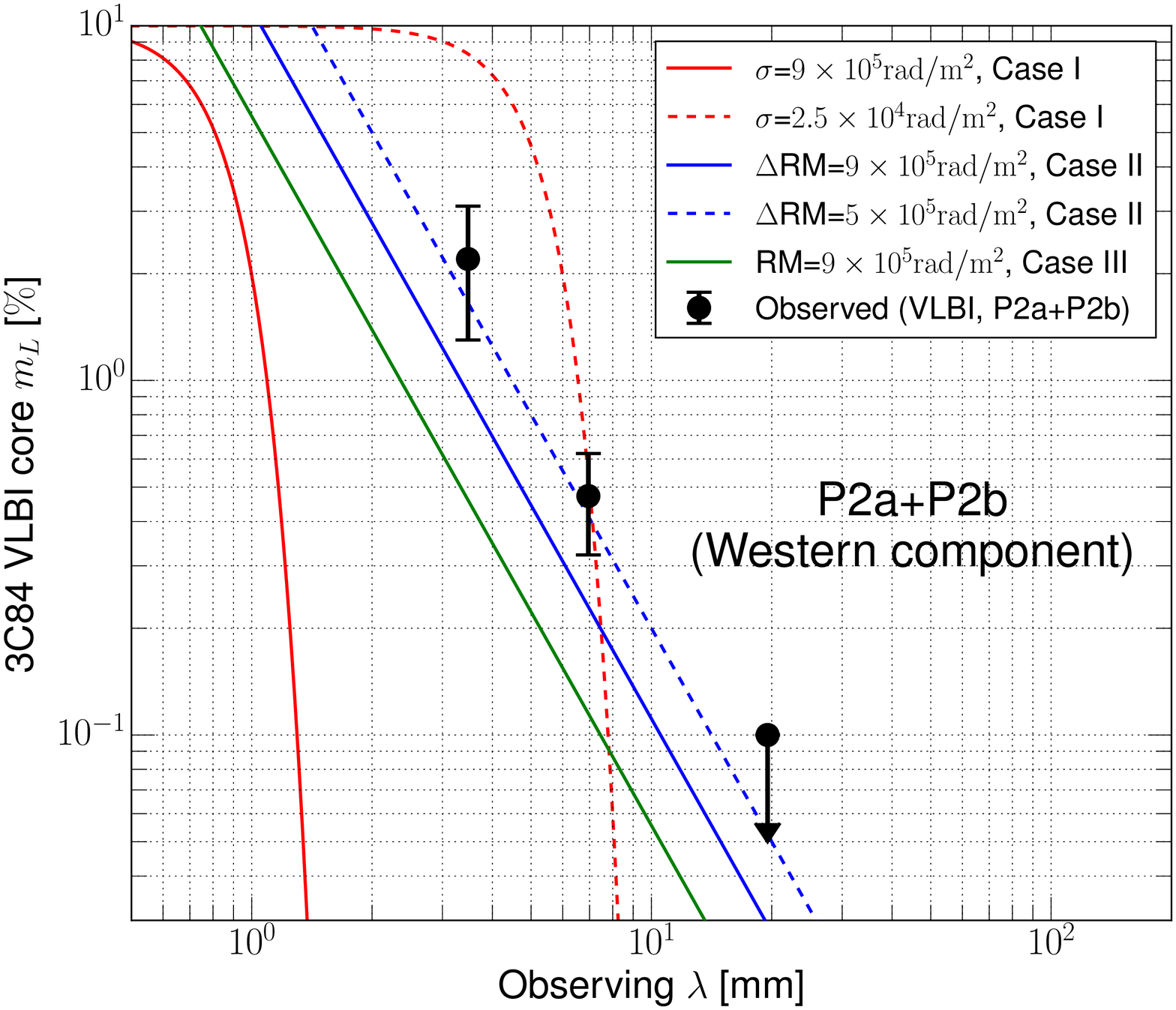}
\includegraphics[width=0.45\textwidth]{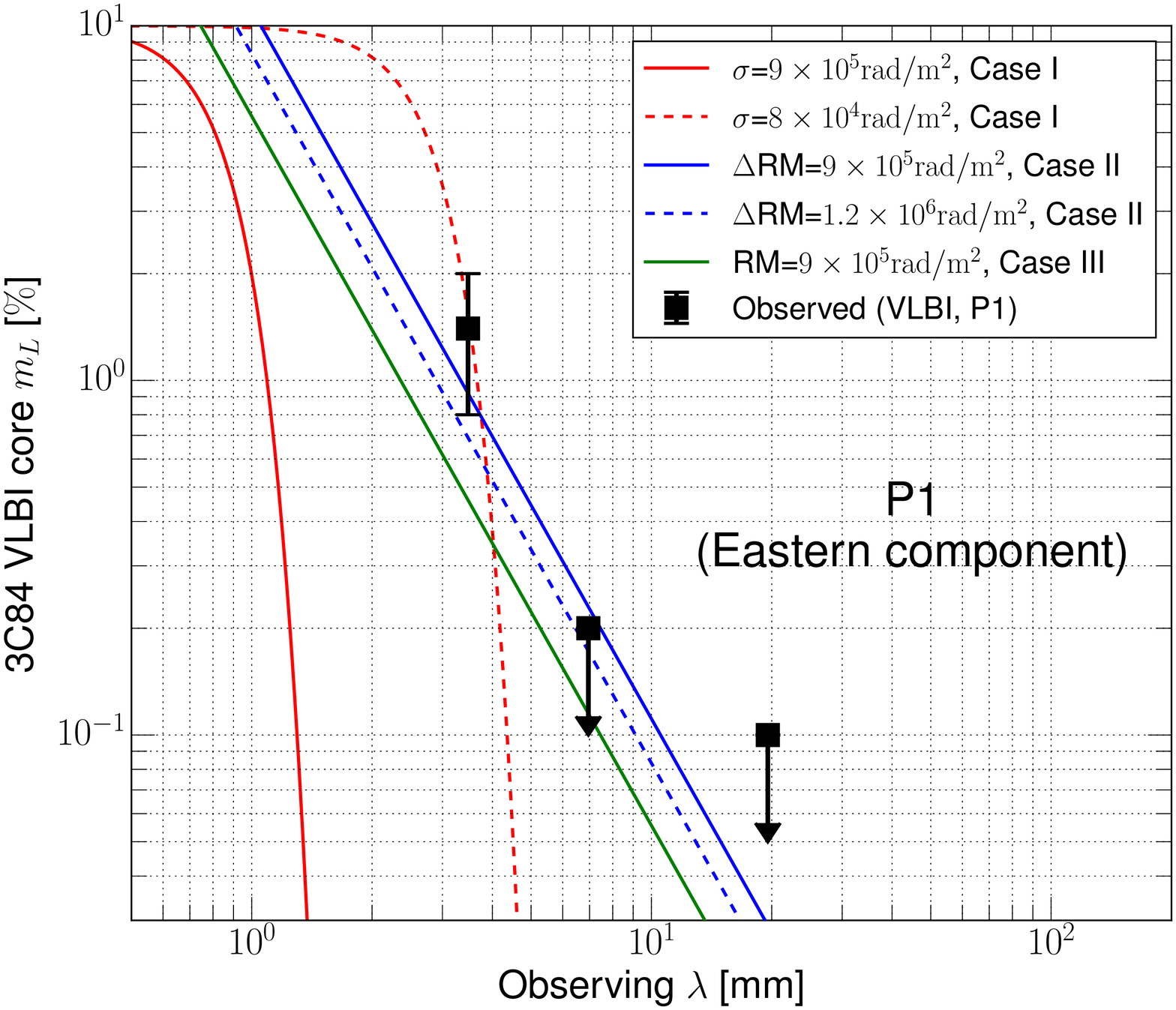}
\caption{
Upper limits for $m_{L}$ for external (Case I and II) and internal (Case III) Faraday depolarization 
and the VLBI polarization from observations.
Arrows indicate upper limit from observations. 
The top and bottom panels are for the P2 and P1 components, respectively.
See the text for the details.
}
\label{fig:rm_ml}
\end{figure}

In Fig. \ref{fig:rm_ml} we show the upper limits for the linear polarization degree calculated for the three depolarization models.
The plots reveal several important points.
In Case I (red lines), the RM dispersion $\sigma$ is as large as $9\times10^{5}~{\rm rad/m^{2}}$ for $\epsilon\sim1$.
This is apparently problematic for both P2 and P1 because of the strong depolarization (solid red lines).
We find that $\sigma\sim2.5\times10^{4} {\rm rad/m^{2}}$ is required for P2
in order to explain the observed linear polarization at 43\,GHz (dashed red line).
An even smaller $\sigma$ would be required 
if the intrinsic degree of polarization $m_{0}$ is smaller than 10\%
for a more disordered magnetic field.
A $\sigma$ which is two orders of magnitude smaller than the RM (i.e., $\epsilon\lesssim(2.5\times10^{4})/(9\times10^{5})\sim0.03$)
indicates a highly uniform RM distribution in the external screen, which requires a well ordered large-scale magnetic field.
Similar results are obtained for P1 ($\sigma$ is at most $\sim8\times10^{4}\rm rad/m^{2}$ and $\epsilon\lesssim(8\times10^{4})/(9\times10^{5})\sim0.09$).
In contrast, Case II (blue lines) seems to explain the observed linear polarization in the core much better.
We find that for P2 and P1 rotation measure gradients of $\Delta{\rm RM}\sim5\times10^{5}~{\rm rad/m^{2}}$ and
$\Delta{\rm RM}\sim1.2\times10^{6}~{\rm rad/m^{2}}$ could explain the polarization at $\ge43$\,GHz, respectively (dashed blue lines).

In Case III (green solid lines), the depolarization exceeds the values that we infer from our data, so we regard Case III as less likely and do not discuss it further.

Therefore, a smooth RM distribution in the external screen appears to be the most plausible scenario.
We now proceed to explore the implications of the model cases I and II, trying to distinguish between them.

\subsubsection{Case I : Accretion flow as the foreground screen}\label{subsec:rm_external}

In the vicinity of the central black hole and depending on its covering factor, a radiatively inefficient accretion flow RIAF (e.g., \citealt{yuan14})
could be regarded as a good candidate for a Faraday screen.
In such a flow, magneto-rotational instabilities and magnetohydrodynamic (MHD) turbulence can introduce a certain degree of inhomogenity in the matter and magnetic field distribution
\citep{balbus91,balbus98}.
Recently, \cite{johnson15} measured the linear polarization in Sgr A* with VLBI at 230\,GHz and suggested 
a short coherence length of the magnetic field orientation, of order of $\sim6R_{s}$. The linear polarization appears
more random over larger scales. Adopting this also for the accretion flow of 3C\,84 leads to the
expectation that the Faraday depth across the screen is comparable to the characteristic Faraday depth
(i.e., $\sigma$ and/or $\Delta{\rm RM}\sim{\rm RM}$).
However, this is in contrast to our finding (see Fig. \ref{fig:rm_ml}).
Certainly, our modeling simplifies a number of details in the disk-jet system.
However, detailed general relativistic MHD simulations of such an accretion flow also suggest
very strong depolarization of the jet by the accretion flow (e.g., \citealt{moscibrodzka17}).
Hence, the accretion flow appears to us as a less preferred candidate for the Faraday screen.

Another problem in the association of the large RM in 3C\,84 with the accretion flow was pointed out by \cite{plambeck14}.
The authors modeled a spherical accretion flow in 3C\,84 using  
a highly ordered, radial magnetic field, in order to relate the observed RM$\sim9\times10^{5}~\rm rad/m^{2}$ 
with the expected mass accretion rate.
For realistic estimates of the magnetic field strength and the electron density in the accretion flow,
the authors derive a higher RM, so that the observed RM appears to be too small.
Therefore, the authors concluded that either 
(i) the magnetic field strength is much weaker than the equipartition value,
(ii) the magnetic field is highly disordered, and/or 
(iii) the accretion flow is rather disk-like and the line of sight does not pass through the accretion flow.
Indeed, an oblate disk-like geometry of the accretion flow would provide a suitable explanation 
because it explains the observational findings without fine-tuning the intrinsic physical properties of the accretion flow.

If the observed linear polarization is marginally affected by the disk-like accretion flow because of the geometry,
the disk height $H$ to disk radius $R$ ratio $H/R$ should be
$\tan^{-1}(H/R)\lesssim(90^{\circ}-\theta)$ or $H/R\lesssim \tan(90^{\circ}-\theta)$ where $\theta$ is the jet viewing angle.
If we adopt $\theta\sim30^{\circ}$ ($\sim60^{\circ}$) from previous observations \citep{walker94,fujita17}, we find $H/R\lesssim1.7$ ($\lesssim0.6$). 
If the thick accretion flow is threaded by a strong poloidal magnetic field which can compress the disk vertically (e.g., \citealt{mckinney12}),
an even smaller $H/R$ ratio may be physically possible.

\subsubsection{Case II : Faraday rotation \& depolarization due to the transverse jet stratification}\label{subsec:rm_internal}

Alternatively, a mildly-relativistic sheath surrounding the relativistic jet may provide the required external screen.
A wide and collimated jet can be formed by the inner accretion disk 
or directly by spinning central black hole
\citep{bp82,bz77}.
The rotation of the central engine leads to the development of 
a stratified and twisted magnetic field topology (e.g., \citealt{tchekhovskoy15}).
The boundary layer (or sheath) of such jets may provide the uniform external Faraday screen in Case II.
Evidence for an ordered magnetic field configuration in jets comes from a plethora of polarization VLBI jet observations and from 
observations of transverse RM gradients on pc scales (e.g., \citealt{zavala05,hovatta12,gabuzda17}).
The magnetic field is expected to be more ordered in the inner jet region when the jet launching region becomes magnetically dominated 
(e.g., \citealt{zamaninasab14,martividal16}).

\begin{figure}[t!]
\centering
\includegraphics[width=0.45\textwidth]{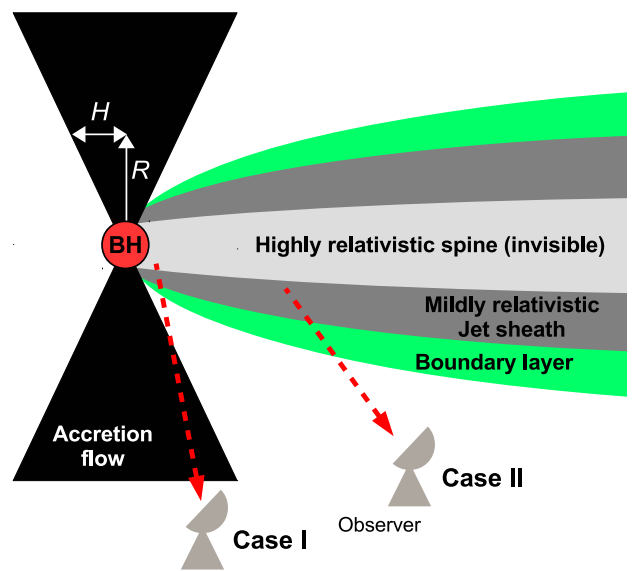}
\caption{
Illustration of the geometry discussed throughout Sect. \ref{subsec:rm_external} and \ref{subsec:rm_internal}.
Note that the figure is not drawn to scale.
}
\label{fig:illustration}
\end{figure}

High angular resolution VLBI observations of the inner jet of 3C\,84 show a limb-brightened morphology
(see Fig. \ref{fig:gmva_inner} and also \citealt{giovannini18}).
A limb-brightened jet most likely consists of at least two different jet components with different speeds, 
electron densities and/or magnetic field strengths (spine-sheath geometry; e.g., \citealt{pelletier89,komissarov90}).
Mildly relativistic electrons in the boundary layer of the jet, possibly mixed with thermal particles, 
will then cause Faraday rotation and conversion
\citep{sokoloff98,perlman99,pushkarev05,porth11,macdonald15,pasetto16,lico17}.
In Figure  \ref{fig:illustration} we show a sketch of a stratified jet for illustration.

\subsection{Estimation of the jet magnetic field strength and the electron density}\label{subsec:ssa_eq_b}

In the following we assume that the boundary layer of the jet is the Faraday screen and is responsible for the observed RM.
We estimate the jet electron number density adopting the jet magnetic field strength from the synchrotron self-absorption theory.
In convenient units, the observed RM can be written as follows:
\begin{equation}
\frac{\rm RM}{\rm rad/m^{2}} = 8.1\times10^{5} \int_{\rm source}^{\rm telescope} 
\left( \frac{n_{e}}{\rm cm^{-3}} \right)
\left( \frac{B_{||}}{\rm G} \right)
\left( \frac{dl}{\rm pc} \right)
\label{eq:RM_eq}
\end{equation}
where 
$n_{e}$ is the number density of the thermal electrons between the source and the telescope,
$B_{||}$ is the line-of-sight component of the magnetic field,
and $dl$ is the path length through the plasma from the source toward the observer.

We calculate the synchrotron self-absorption (SSA) magnetic field strength $B_{SSA}$ in the core following \cite{marscher83};
\begin{equation}
B_{SSA} = 10^{-5}b(\alpha) \frac{\theta^{4}_{m}\nu^{5}_{m}\delta}{S^{2}_{m}(1+z)} \quad \rm Gauss
\label{eq:b_ssa}
\end{equation}
where
$b(\alpha)$ is a constant tabulated in \cite{marscher83} 
as a function of the spectral index, 
$S_{m}$ is the flux density (in Jy) at the turn-over frequency $\nu_{m}$ (in GHz),
$\theta_{m}$ is the source angular size (in mas) at $\nu_{m}$,
and $\delta$ is the Doppler factor.
We note that the exact value of $\nu_{m}$ is not accurately determined by our data (see the spectrum in Fig. \ref{fig:core_sed}).
However, this spectrum and recent VLBI observations of 3C\,84 at 86\,GHz and 129\,GHz (\citealt{hodgson18})
suggest a spectral turnover near 86\,GHz. Therefore, we adopt $\nu_{m}=86$\,GHz for the calculation of $B_{SSA}$.
For the optically thin spectrum we use $\alpha=-0.5$, which fixes $b(\alpha)=3.2$.
Assuming $\nu_{m}=86$\,GHz, $S_{m}=(5.6\pm1.7)$\,Jy, and $\theta_{m}=1.8\times(0.14\pm0.01)$\,mas\footnote{
The factor 1.8 corrects for the geometry (\citealt{marscher83}).},
we obtain $B_{SSA}\sim(19\pm13)\times\delta$\,G.  

For the jet viewing angle $\theta\sim30^{\circ}-60^{\circ}$ and the apparent jet speed $\sim0.1c$
near the core, the Doppler factor is $\delta\sim1.1$. 
With this, we obtain $B_{SSA}\sim(21\pm14)$\,G.
We note that \cite{abdo09} and \cite{aleksic14} suggest a rather large Doppler factor of $\delta\sim2-4$
based on the theoretical modeling of the synchrotron self-Compton process in the jet.
In this case, the resulting magnetic field strength would be $\sim(2-4)$ times larger.

We take $B_{SSA}$ as the average magnetic field strength in the jet and
calculate the average electron number density $n_{e}$ by

\begin{equation}
\frac{n_{e}}{\rm cm^{-3}} \sim 1.2\times10^{-6}
\left( \frac{{\rm RM} }{\rm rad/m^{2}} \right)
\left( \frac{B_{SSA}}{\rm G} \right)^{-1}
\left( \frac{\theta_{m}}{\rm pc} \right)^{-1}
\label{eq:ne}
\end{equation}

\noindent
where we assume that the size $\theta_{m}$, is a lower limit to the approximate pathlength for the 
jet medium because a longer pathlength is expected for a deprojected line-of-sight.
For ${\rm RM}=9\times10^{5}{\rm rad/m^{2}}$, we obtain $n_{e} \leq (0.6\pm0.4)~{\rm cm^{-3}}$.
The density will be even lower if the Doppler factor is larger and the magnetic field strength is higher.
In any case, the density is an order of magnitude lower than an
ambient gas density estimate of  $n_{e}\sim8~{\rm cm^{-3}}$ obtained by \cite{fujita17} for the central parsecs.
A low jet density is not implausible and is consistent with jet collimation by a 
denser ambient medium (e.g., \citealt{nagai14,giovannini18}).

We note that our calculations make simplified assumptions with regards to 
the magnetic field strength, the path length, and the jet geometry. 
Future more detailed calculations using numerical radiative transfer across all Stokes parameters  
(e.g., \citealt{macdonald16}) 
should be able to constrain these parameters more precisely.
Also, the exact value of the turn-over frequency on this spatial scale is still quite uncertain.
Quasi-simultaneous global mm-VLBI observations with the GMVA and the Event Horizon Telescope (e.g., \citealt{doeleman12,lu18})
will help to determine the spectral properties of the innermost jet and central engine.

\section{Summary and Conclusions}\label{sec:conclusion}

In this paper, we presented a study of the polarization properties of the radio galaxy 3C\,84 
based on the results from polarimetric VLBI observations of the source at mm-wavelengths.
We summarize our findings and main conclusions as follows:

\begin{enumerate}

\item
We found asymmetrically distributed and polarized emission in the VLBI core region of 3C\,84 at 86\,GHz using GMVA observations. 
The east-west oriented and linearly polarized structure consists of two polarized components
(P1 and P2), which are separated by $\sim 0.2$\,mas (corresponding to $\sim0.07\textrm{pc}\sim812R_{s}$). 
The fractional linear polarization $m_{L}$ of P1 and P2 at 86\,GHz is $\sim2$\% with the VLBA 43\,GHz beam.

\item
Additional close-in-time VLBA 43\,GHz polarization images also reveal the presence of polarized emission ($m_{L}\sim0.4-0.5$\%).
The polarized structure is also time-variable. The moving direction of the variable polarized emission feature, 
which is almost perpendicular to the milli-arsecond jet,
suggests that the jet base may be edge-brightened in polarized light.
However, other effects such as ejection of a new VLBI component or opacity effects across the jet cannot be excluded.

\item 

The total intensity spectrum of 3C\,84 is inverted up to at least 86\,GHz
and the linear polarization $m_{L}$ increases with frequency following
a power-law $m_{L}\propto\nu^{0.77}$.
A likely explanation invokes a decreasing synchrotron opacity
and lower  Faraday depolarization at higher frequencies.

\item 
The combination of quasi-simultaneous EVPA measurements at 43 GHz and 86 GHz, and adding near in time
EVPA measurements from ALMA at higher frequencies, reveals a high rotation measure in the VLBI core region of 
RM $\sim2\times10^{5}~{\rm rad/m^{2}}$ at $\gtrsim43$\,GHz in May 2015. This is 
comparable to the RM measurements at 230\,GHz on larger angular scales \citep{plambeck14}.
\item
A significant rotation of the EVPA is also observed within the 43 GHz VLBA observing band and also suggests the presence of Faraday rotation.
Repeated measurements of the RM within the 43\,GHz VLBA observing band are consistent with the high
RM in the VLBI core region at three epochs, but also suggest that the RM has changed its sign at one epoch. This 
could be related to the onset of a small total intensity flare in the VLBI core in 2015. 

\item
In order to explain the degree of linear polarization by Faraday depolarization,
an external screen with either a small RM dispersion or
only a smoothly varying uniform RM distribution is required.

\item

The data would support a tentative association of the Faraday screen with the accretion flow if 
(i) the accretion flow is thick ($H/R \geq 1.7$ for the jet viewing angle $\theta=30^{\circ}$) and more importantly 
(ii) the magnetic field is highly ordered.
However, the latter appears to be in contradiction with the results of previous studies (e.g., \citealt{plambeck14}).
Instead, a stratified jet with a boundary layer containing an ordered magnetic field configuration could also provide 
a good explanation for the large RM, the frequency dependence of the depolarization, and possibly also for the RM variability.
The edge-brightened jet morphology observed at 22\,GHz \citep{giovannini18} would also support this interpretation. 

\item
From synchrotron self-absorption theory, we calculate a magnetic field strength in the VLBI core region of 3C\,84 of 
$B_{SSA}=(21\pm14)$\,G,
adopting an only mild relativistic beaming with a Doppler factor of $\delta\sim1.1$ from jet kinematics.
In this case a jet electron density of $n_{e}\leq(0.6\pm0.4)~{\rm cm^{-3}}$ is required to explain the observed high RM.
The field strength would be larger by factor $\sim2-4$ if a higher Doppler factor of $\delta\sim2-4$ is used,
and the electron density will be correspondingly lower. 
In both cases the jet electron density appears to be at least an order of magnitude lower than
the density of the ambient gas.

\end{enumerate}

Overall, our study suggests that either the accretion flow or a transversely stratified jet with boundary layers can 
play an important role in the generation of the polarized mm-wave emission in the milli-parsec scale VLBI structure.
Future millimeter VLBI imaging, performed quasi-simultaneously at different frequencies
and including VLBI observations above 86\,GHz, will help to better understand the complicated nature of the 
polarization in 3C\,84.
Also, a more detailed theoretical modeling of the linear and circular polarization spectrum, and complemented by 
numerical simulations, would be very beneficial.
Finally, we note that the detection of significant linear polarization at the jet base in 3C\,84 may
shed light on the properties of the Faraday screen in other radio galaxies (e.g., M\,87; \citealt{kim18}). 
Further studies should prove fruitful.

\begin{acknowledgements}
%
%
We thank the anonymous referee for valuable comments and suggestions, which greatly helped to improve the paper.
%
%
We thank Carolina Casadio and Ioannis Myserlis for fruitful discussions and comments.
J.-Y.K. is supported for this research by the International Max-Planck Research School (IMPRS) 
for Astronomy and Astrophysics at the University of Bonn and Cologne. 
%
%
I.A. acknowledges support by a Ram\'on y Cajal grant of the Ministerio de Econom\'ia, Industria y Competitividad (MINECO) of Spain, 
and by and additional MINECO grant with reference AYA2016--80889--P.
%
%
R.-S. L. is supported by the National Youth Thousand Talents Program of China and by the Max-Planck Partner Group.
%
This research has made use of data obtained with the Global Millimeter VLBI Array (GMVA), 
which consists of telescopes operated by the MPIfR, IRAM, Onsala, Metsahovi, Yebes, 
the Korean VLBI Network, the Green Bank Observatory and the Long Baseline Observatory. 
The VLBA is an instrument of the Long Baseline Observatory,
which is a facility of the National Science Foundation operated by Associated Universities, Inc. 
The data were correlated at the correlator of the MPIfR in Bonn, Germany.
%
This work made use of the Swinburne University of Technology software correlator, 
developed as part of the Australian Major National Research Facilities Programme and operated under licence.
%
%
This study makes use of 43 GHz VLBA data from the VLBA-BU Blazar Monitoring Program (VLBA-BU-BLAZAR;
http://www.bu.edu/blazars/VLBAproject.html), funded by NASA through the Fermi Guest Investigator Program. 
%
%
This research has made use of data from the MOJAVE database that is maintained by the MOJAVE team \citep{lister09}.
%
%
This paper makes use of the following ALMA data: ADS/JAO.ALMA\#2011.0.00001.CAL. 
ALMA is a partnership of 
ESO (representing its member states), NSF (USA) and NINS (Japan), together with NRC (Canada), MOST and ASIAA (Taiwan), 
and KASI (Republic of Korea), in cooperation with the Republic of Chile. 
The Joint ALMA Observatory is operated by ESO, AUI/NRAO and NAOJ.
%
%
This research has made use of data from the OVRO 40\,m monitoring program \citep{richards11}.
The OVRO 40\,M Telescope Fermi Blazar Monitoring Program is supported by 
NASA under awards NNX08AW31G and NNX11A043G, and by the NSF under awards AST-0808050 and AST-1109911. 
\end{acknowledgements}

%
%

\bibliographystyle{aa}
\bibliography{3c84.bib}

\begin{appendix}

\section{On the flux density calibration of 3C\,84 using 86\,GHz GMVA data in May 2015}

\begin{figure}[t]
\centering
\includegraphics[width=0.45\textwidth]{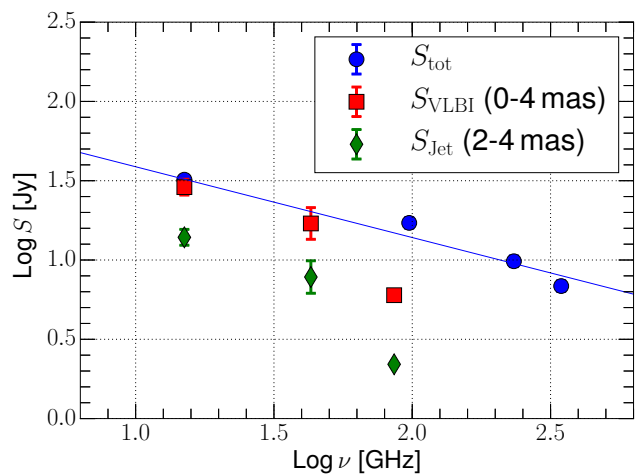}
\caption{
Quasi-simultaneous total and VLBI-scale flux densities of 3C\,84 in May 2015 displayed in log-log scale.
The blue solid line shows the power-law fitted to the total flux measurements.
We omit the error bars for the GMVA 86\,GHz data points. 
}
\label{fig:appendix_1}
\end{figure}

Here we discuss a method to improve the accuracy of the absolute flux density calibration of 3C\,84 at 86\,GHz based on the available
a-priori GMVA amplitude calibration, which is not accurate enough (see also \citealt{koyama16}).
We compared the multi-frequency VLBI flux densities of 3C\,84 (Sect. \ref{subsec:gmva} and \ref{subsec:low_vlbi})
with the total flux densities measured by other instruments.
At $\nu=15$\,GHz, the Owens Valley Radio Observatory (OVRO; \citealt{richards11}) monitoring program\footnote{http://www.astro.caltech.edu/ovroblazars/}
provided near-in-time flux density measurements of $S=31.8\pm0.5$\,Jy and $S=32.3\pm0.1$\,Jy for 3C\,84 on 03 Apr 2015 and 27 Jun 2015, respectively.
We averaged the two fluxes in order to estimate the total flux in May 2015 and 
took half of their difference as the associated error, finding $S=32.1\pm0.3$\,Jy.
At $\nu\gtrsim86$\,GHz, we refer to Sect. \ref{subsec:alma} for the ALMA flux density measurements.
In Fig. \ref{fig:appendix_1} we show 
(i) the VLBI-scale flux density of the source within $\sim4$\,mas from the core ($S_{\rm VLBI}$),
(ii) the VLBI-scale flux density only for the extended jet ($\sim2-4$\,mas from the core; $S_{\rm Jet}$), and
(iii) the total (i.e., spatially unresolved) flux density measured by the OVRO and the ALMA observations ($S_{\rm tot}$).
The GMVA flux densities were taken from only the self-calibrated data, based on the a-priori calibration.
We fitted a power-law model to the total flux density (i.e., $S_{\rm VLBI}\propto \nu^{+\alpha}$) 
in order to estimate the VLBI flux density $S_{\rm VLBI}$ at 43 and 86\,GHz.
The total flux density spectrum is fitted by $S_{\rm tot}\propto \nu^{-(0.45\pm0.04)}$.
From this we estimate $S_{\rm tot}$ at 43\,GHz and 86\,GHz
and calculated the VLBI-to-total flux density ratio $S_{\rm VLBI}/S_{\rm tot}$ (i.e., the compactness).
At 15\,GHz and 43\,GHz, the flux ratio is comparable ($\sim0.90$ and $\sim0.84$ at 15\,GHz and 43\,GHz, respectively).
However at 86\,GHz, it is much lower ($S_{\rm VLBI}/S_{\rm tot}\sim0.40$).
The comparison of the spectral index of the extended jet emission  yields a spectral
index of $\alpha_{\rm 15-43\,GHz}\sim-0.5$, while the spectral index from 43\,GHz to 86\,GHz is much steeper;
$\alpha_{\rm 43-86\,GHz}\sim-1.8$. Such a steep spectrum is unlikely, and is inconsistent with ALMA measurements
between 98\,GHz and 344\,GHz. We therefore take this as another indication for a too low flux densities at 86\,GHz,
%


To check for an overall scaling problem of the GMVA amplitude calibration
we examined the compactness ratio $S_{\rm VLBI}/S_{\rm tot}$ also for other calibrators at 86\,GHz (3C\,454.3, BLLac, CTA\,102, and OJ\,287), for which
we find -- applying a similar procedure as described above -- compactness ratios
of typically 0.9 at 15\,GHz and 43\,GHz, but $\sim 0.5$ at 86\,GHz.
In the following we assume that the true compactness $S_{\rm VLBI}/S_{\rm tot}$ at 15, 43, and 86\,GHz 
is comparable. Then the visibility amplitudes of the GMVA data of 3C\,84 can be upscaled by a factor $g\sim2.1-2.3$.
Confidence for this number also come from similar factors obtained for the other calibrators.
With this correction factor a much more realistic value of the (optically thin) jet spectral index $\alpha_{\rm 43-86\,GHz}\sim-0.8$ 
is obtained.
The corrected total flux density ($\sim12$\,Jy) is now also in good agreement with other 
near-in-time VLBI flux density measurements at 86\,GHz reported in literature (e.g., $S \sim(10-15)$\,Jy; see \citealt{hodgson18}).

In the following we discuss the uncertainty $\Delta g$ of the scaling value $g$, which depends on the validity of our assumptions, 
and we could alternatively assume the same spectral index from 15\,GHz to 86\,GHz for $S_{\rm Jet}$. 
Assuming a constant spectral index $\alpha_{\rm 15-86\,GHz}\sim-0.5$ results in a slightly larger correction factor of $g\sim2.5$.
The range of the $g\sim2.1 - 2.5$ suggests that its uncertainty is $\Delta g \sim0.4$, corresponding to a $\sim20$\% of error in the absolute flux density.
It would be overly optimistic to claim an error smaller than $\sim20$\% without having identified the origin of the amplitude scaling error.
Therefore, we adopted a slightly larger value of $\sim30$\% for the systematic uncertainty in the measured VLBI flux densities (for this particular epoch).

\section{ Robustness of the polarization imaging at 86~GHz }\label{appendix:imaging_test}

\begin{figure}[!ht]
\centering
\includegraphics[width=0.43\textwidth]{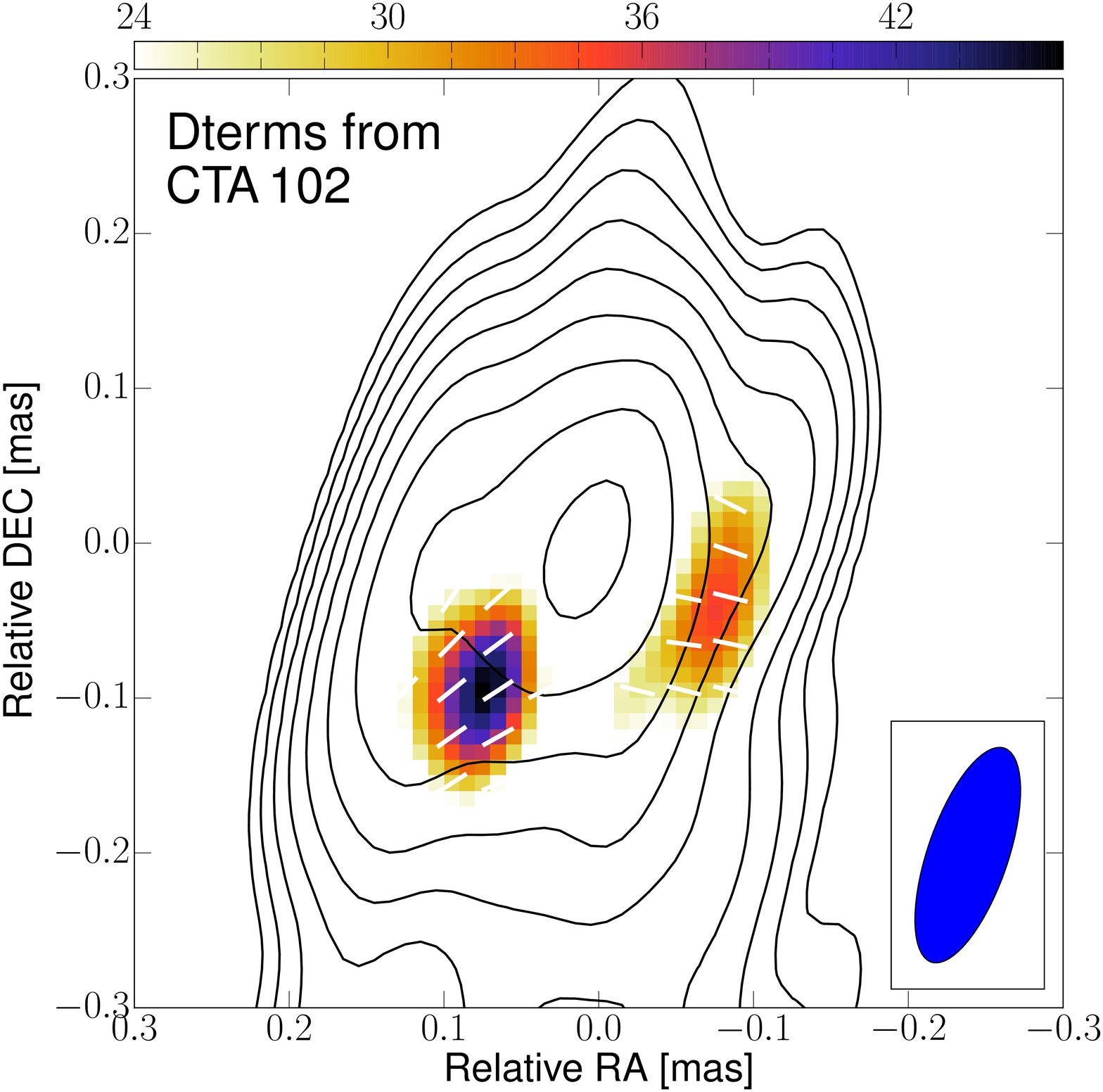}
\includegraphics[width=0.43\textwidth]{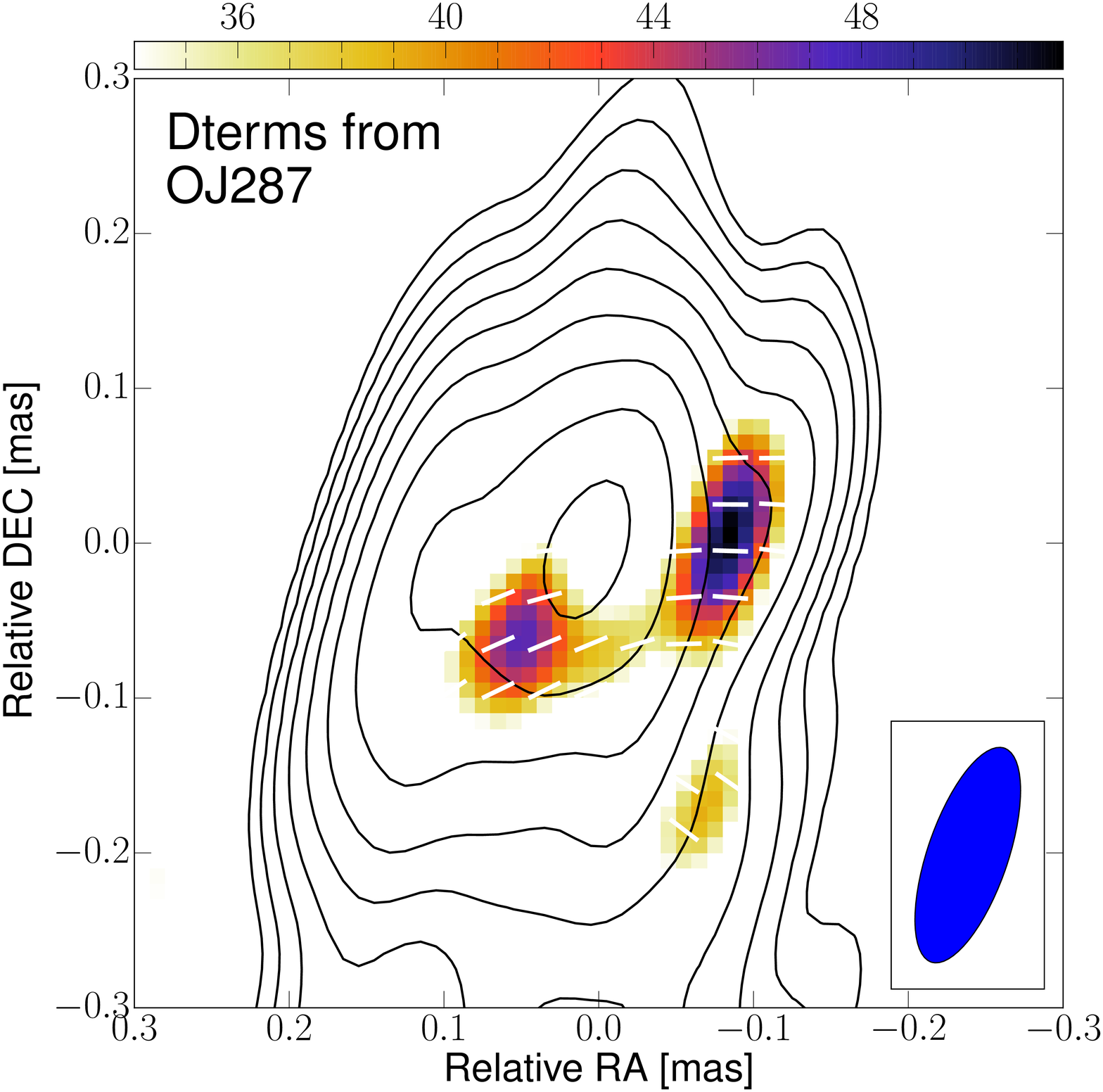}
\caption{
Linear polarization maps of the nuclear region of 3C\,84 at 86\,GHz
obtained with different polarization calibrations. 
The two maps were reconstructed using the D-terms obtained from 
CTA\,102 (upper panel) and from OJ\,287 (lower panel), respectively.
The minimum contour level is 7\,mJy/beam.
       }
\label{fig:dterm_cal}
\end{figure}

In order to 
illustrate the impact of slightly different D-terms on the polarization image of the GMVA data, 
we made different linear polarization images of 3C\,84 using 
D-terms determined from the individual calibrators.
Two examples are shown in Fig. \ref{fig:dterm_cal}.
We find that two central polarization features are reproduced in the nuclear region at 
similar locations with similar EVPAs.
We note that the structure of the jet of OJ\,287 is 
substantially different from that of CTA\,102
(e.g., \citealt{agudo12,casadio15,hodgson17}). 
Despite this, the two polarization maps shown in Fig. \ref{fig:dterm_cal} look similar, 
which provides confidence of the polarization calibration and imaging of 3C\,84.
Nevertheless, we consider that the polarization image in the main text, which 
is obtained with the averaged D-terms, is of better quality than those in Fig. \ref{fig:dterm_cal} 
for the reasons explained in the main text.

\section{Multi-epoch VLBA 43\,GHz data}\label{appendix:43ghz}

\begin{figure*}[!ht]
\centering
\includegraphics[width=0.45\textwidth]{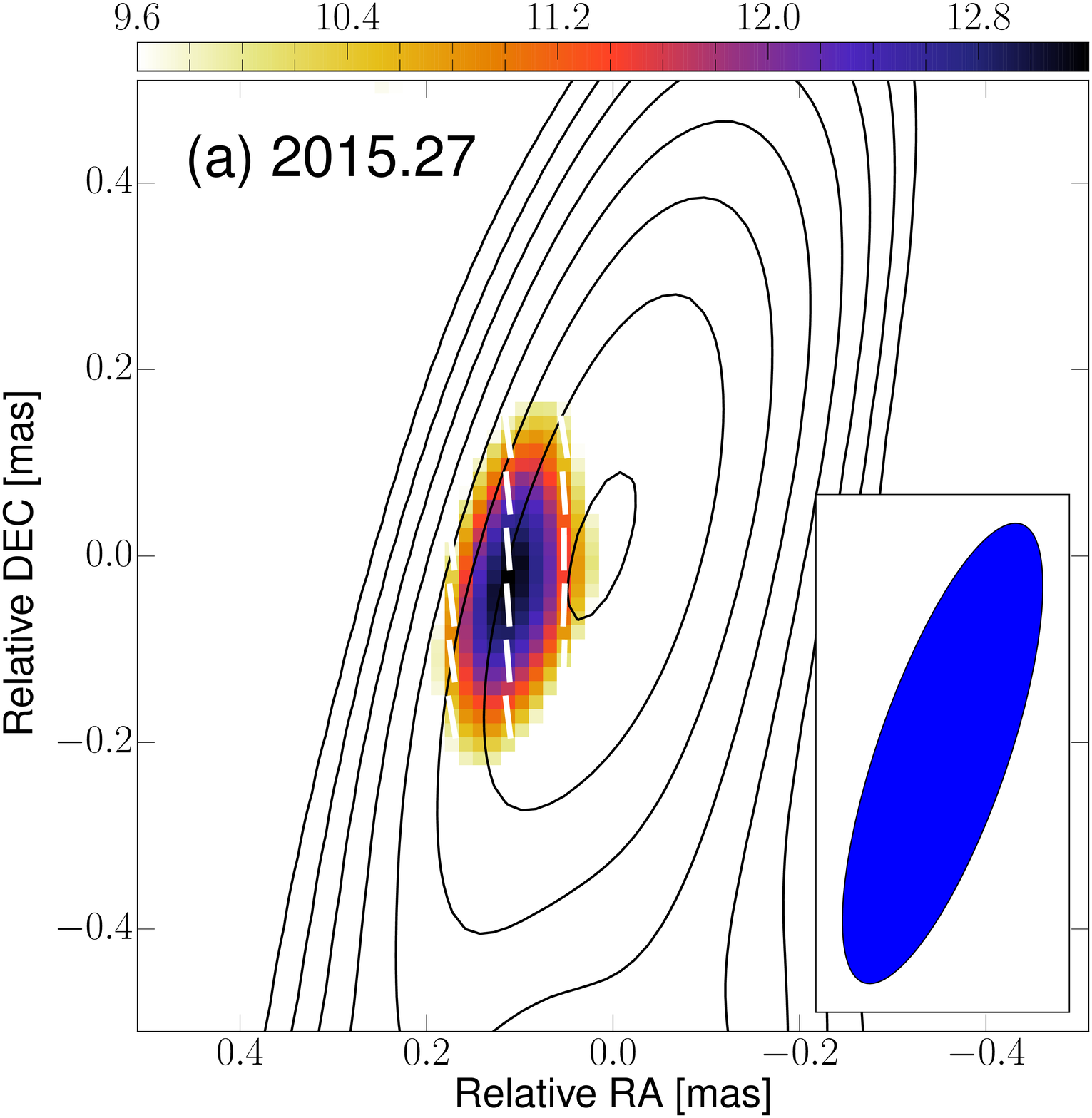}
\includegraphics[width=0.45\textwidth]{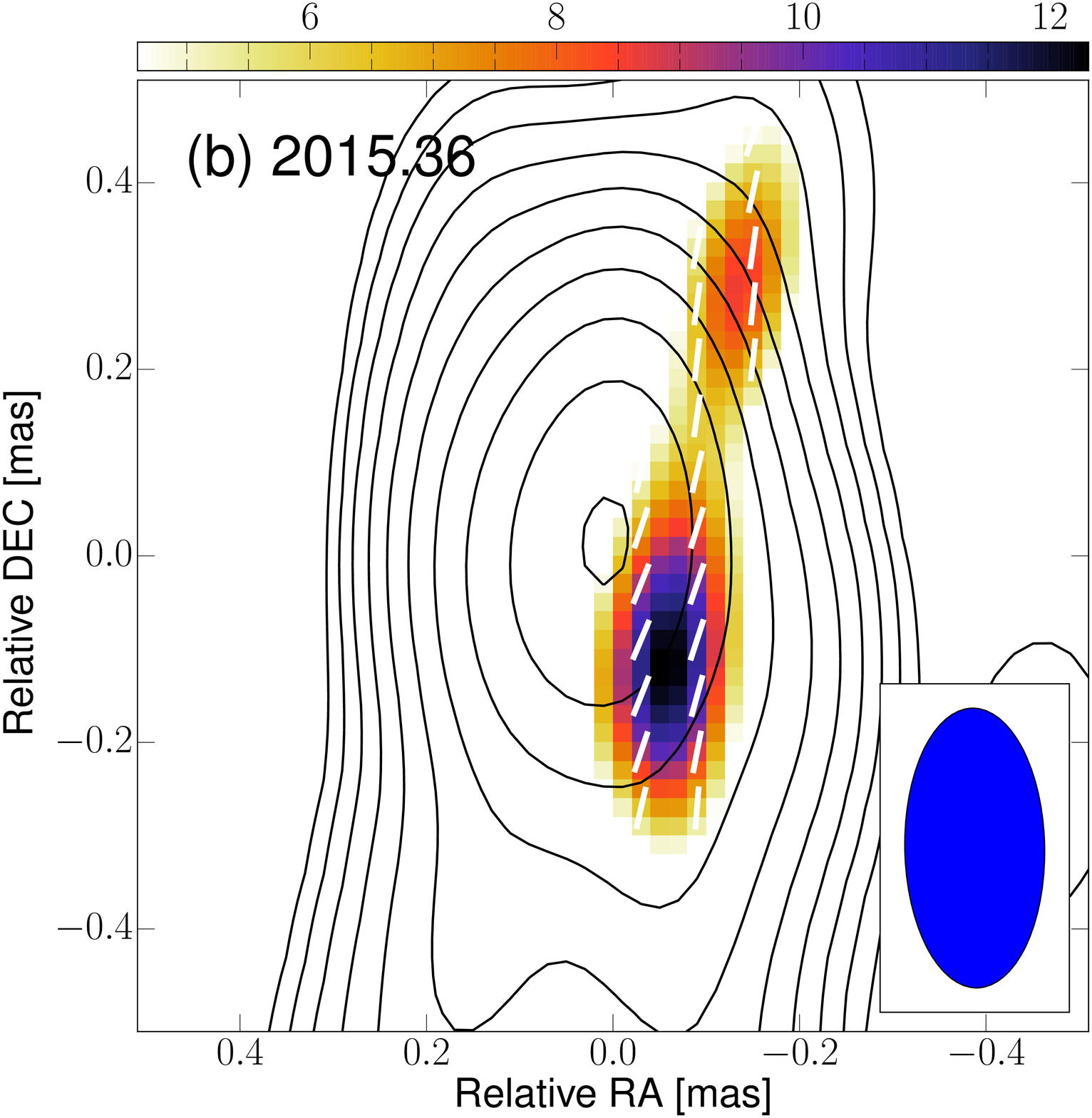}
\includegraphics[width=0.45\textwidth]{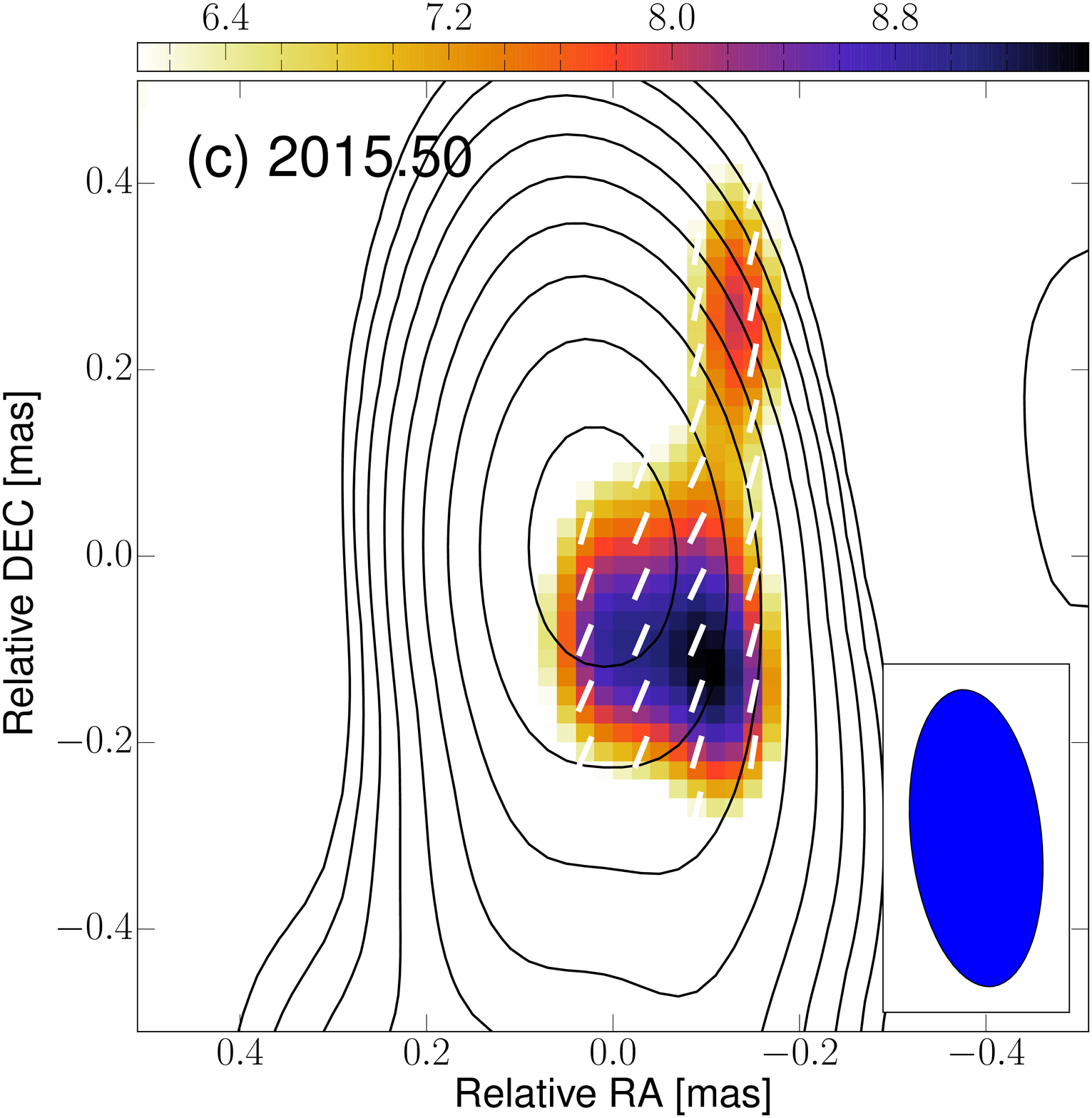}
\includegraphics[width=0.45\textwidth]{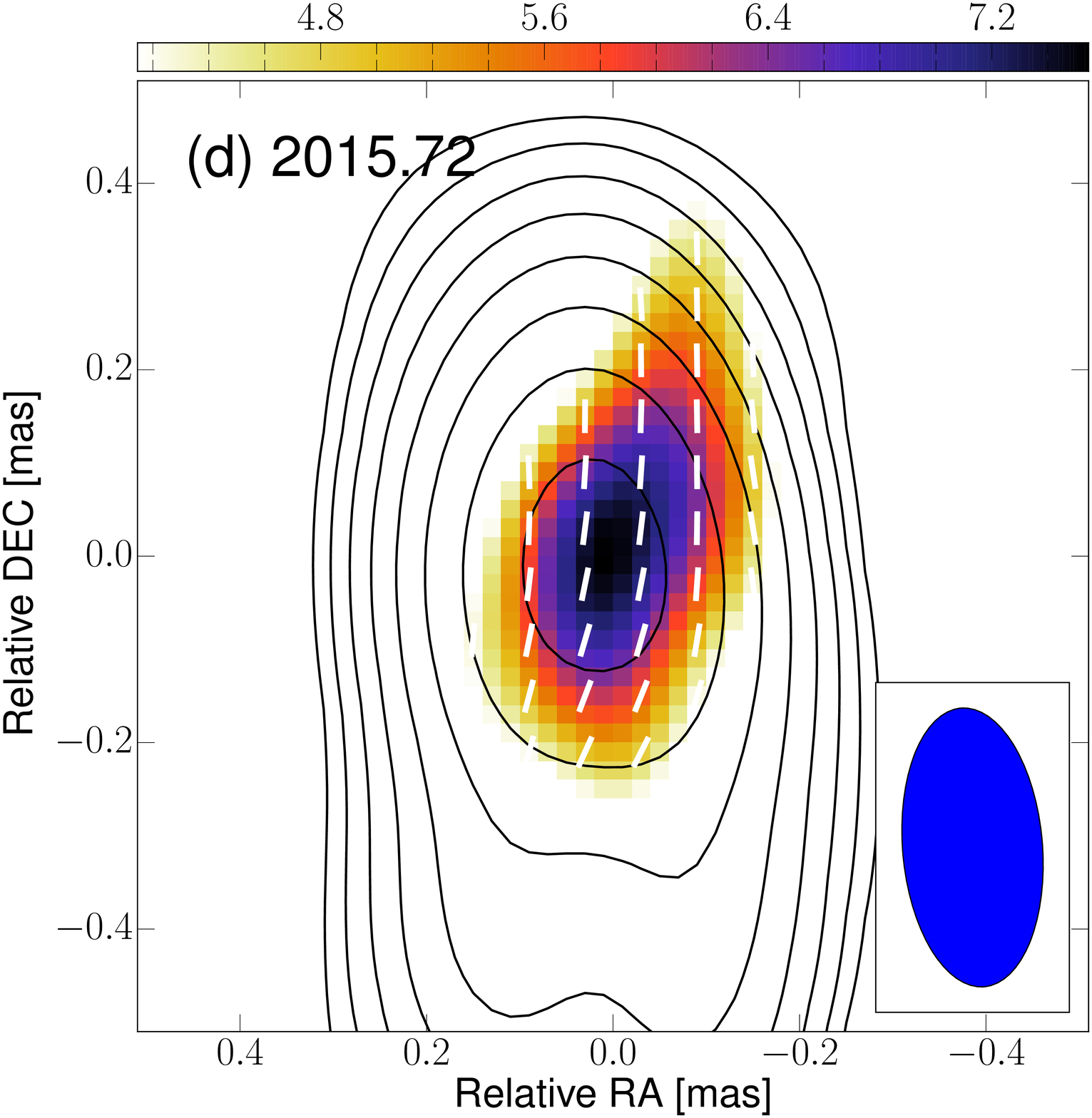}
\caption{
VLBA 43\,GHz polarization images of 3C\,84 zoomed into the core.
Labels in the upper left of each panel denotes the observing epoch.
The contours show the total intensity and start from 25, 7, 10, and 20~mJy/beam 
and increase by factor of 2 for the panels (a), (b), (c), and (d), respectively.
The colorbar indicates the linear polarization intensity in mJy/beam.
       }
\label{fig:bu_pol_maps}
\end{figure*}

In Table \ref{tab:summary_appendix} we show the results of the VLBA 43\,GHz polarization measurements 
in the four different epochs.
We provide the polarization maps of the corresponding epochs obtained from the frequency-averaged visibilities
in Fig. \ref{fig:bu_pol_maps}.
The core flux and the FWHM size during 2015 are given in Table \ref{tab:summary_appendix2}.


\begin{table}[!ht]
\caption{
Properties of the polarized components at 43\,GHz obtained from the 
polarization imaging at four different IFs.
The columns show
(1) the observing epoch,
(2) the observing frequency at each IF, 
(3) the total flux,
(4) the degree of linear polarization, and
(5) the EVPA.
The total flux was consistent across the frequency in each epoch and 
we report the measurement only for a single IF.
}              
\label{tab:summary_appendix}      
\centering                                      
\resizebox{0.5\textwidth}{!}{%
\begin{tabular}{ccccc}          
\hline\hline                        
Epoch & $\nu_{\rm obs}$ & $S_{\nu}$\tablefootmark{a} & $m_{L}$ & $\chi$ \\
 (1) & (2) & (3) & (4) & (5)  \\\relax
 [yyyy/mm/dd] & [GHz] & [Jy] & [\%] & [deg] \\
\hline                                   
2015/04/11& 
  43.0075 &  $1.4\pm0.1$ & $0.73\pm0.23$   & $3\pm10$  \\
& 43.0875 &              & $0.76\pm0.24$   & $2\pm10$  \\
& 43.1515 &              & $0.61\pm0.19$   & $8\pm10$  \\
& 43.2155 &              & $0.32\pm0.10$   & $20\pm10$ \\
2015/05/11& 
  43.0075 &  $2.0\pm0.2$  & $0.44\pm0.14$   & $-(9\pm10)$   \\
& 43.0875 & 	          & $0.49\pm0.16$   & $-(20\pm10)$  \\
& 43.1515 &    		  & $0.53\pm0.17$   & $-(24\pm10)$  \\
& 43.2155 &  		  & $0.43\pm0.13$   & $-(15\pm10)$  \\
2015/07/02& 
  43.0075 &  $3.6\pm0.4$  & $0.37\pm0.12$   & $-(5\pm10)$   \\
& 43.0875 &  		  & $0.40\pm0.13$   & $-(22\pm10)$    \\
& 43.1515 &  		  & $0.67\pm0.21$   & $-(26\pm10)$    \\
& 43.2155 &  		  & $0.47\pm0.15$   & $-(23\pm10)$    \\
2015/09/22& 
  43.0075 &  $2.8\pm0.3$  & $0.28\pm0.09$   & $-(4\pm10)$   \\
& 43.0875 &  		  & $0.37\pm0.10$   & $6\pm10$    \\
& 43.1515 &  		  & $0.33\pm0.11$   & $-(10\pm10)$    \\
& 43.2155 &	          & $0.38\pm0.12$   & $-(15\pm10)$    \\
\hline                                             
\end{tabular}
}
\tablefoot{
\tablefoottext{a}{The time-variable total flux density of the polarized component is due to 
not only the flux variability but also the different positions of the peak of the polarization.}
}
\end{table}

\begin{table}[!ht]
\caption{
The unpolarized core flux (in Jy), the peak (in Jy/beam), and the core FWHM size (in mas) during 2015
measured from the archival VLBA 43\,GHz data sets by the model-fitting analysis. 
A circular beam of 0.3~mas was used for all epochs.
}              
\label{tab:summary_appendix2}      
\centering                                      
\begin{tabular}{cccc}          
\hline\hline                        
Epoch & Flux & Peak & FWHM size \\\relax
[yyyy/mm/dd] & [Jy] & [Jy/beam] & [mas] \\
\hline
2015/02/14 & $3.9\pm0.7$ & $3.6\pm0.5$ & $0.12\pm0.01$ \\
2015/04/11 & $4.5\pm0.6$ & $4.1\pm0.4$ & $0.13\pm0.01$ \\
2015/05/11 & $4.8\pm0.7$ & $4.4\pm0.4$ & $0.13\pm0.01$ \\
2015/06/09 & $5.8\pm0.6$ & $5.2\pm0.4$ & $0.14\pm0.01$ \\
2015/07/02 & $5.8\pm0.6$ & $5.2\pm0.4$ & $0.16\pm0.01$ \\
2015/08/01 & $5.0\pm0.5$ & $4.5\pm0.3$ & $0.16\pm0.01$ \\
2015/09/22 & $5.0\pm0.6$ & $4.4\pm0.4$ & $0.16\pm0.01$ \\
2015/12/05 & $4.1\pm0.5$ & $3.7\pm0.3$ & $0.15\pm0.01$ \\
\hline                                             
\end{tabular}
\end{table}

\end{appendix}

\end{document}